\documentclass[a4paper,fleqn,usenatbib]{mnras}

\usepackage{newtxtext,newtxmath}
\usepackage[T1]{fontenc}
\usepackage{ae,aecompl}

\usepackage[normalem]{ulem}
\usepackage{xfrac}
\usepackage{float}
\usepackage{epsfig}
\usepackage{color}
\usepackage{graphicx}
\usepackage{longtable}
\usepackage{comment}
\usepackage{amsmath}
\usepackage{mathtools}
\usepackage{threeparttablex}
\usepackage[usenames,dvipsnames]{xcolor}

\def\z{redshift}

\def\zpeak{$z_{\mathrm{peak}}$}
\def\zphot{$z_{\mathrm{phot}}$}
\def\zspec{$z_{\mathrm{spec}}$}

\def\mstar{$M^*$}

\def\msol{$M_{\odot}$}

\title[Imprint of Local Environment on the Galaxy SMF]{Glimpsing the Imprint of Local Environment on the Galaxy Stellar Mass Function}

\author[A. R. Tomczak et al.]
{\parbox{\textwidth}{Adam R. Tomczak$^{1}$\thanks{E-mail: artomczak@ucdavis.edu},
Brian C. Lemaux$^{1}$,
Lori M. Lubin$^{1}$,
Roy R. Gal$^{2}$,
Po-Feng Wu$^{2,3}$,
Bradford Holden$^{4}$,
Dale D. Kocevski$^{5}$,
Simona Mei$^{6,7,8}$,
Debora Pelliccia$^{1}$,
Nicholas Rumbaugh$^{9}$,
Lu Shen$^{1}$
}\\\\
$^1$ Department of Physics, University of California, Davis, One Shields Ave., Davis, CA 95616, USA\\
$^2$ University of Hawai'i, Institute for Astronomy, 2680 Woodlawn Drive, HI 96822, USA\\
$^3$ Max-Planck Institut f\"{u}r Astronomie, K\"{o}nigstuhl 17, D-69117 Heidelberg, Germany\\
$^4$ UCO Lick Observatory, Department of Astronomy and Astrophysics, University of California, Santa Cruz, CA, USA\\
$^5$ Department of Physics and Astronomy, Colby College, Waterville, ME 04901, USA\\
$^6$ LERMA, Observatoire de Paris, PSL Research University, CNRS, Sorbonne Universit\'es, UPMC Univ. Paris 06, F-75014 Paris, France\\
$^7$ University of Paris Denis Diderot, University of Paris Sorbonne Cit\'e (PSC), 75205 Paris Cedex 13, France\\
$^8$ Jet Propulsion Laboratory, Cahill Center for Astronomy \& Astrophysics, California Institute of Technology, 4800 Oak Grove Drive, Pasadena, California, USA\\
$^9$ National Center for Supercomputing Applications, University of Illinois, 1205 West Clark St., Urbana, IL 61801, USA\\
}

\date{Accepted: August 2017. Received May 2017}

\pubyear{2017}

\begin{document}
\label{firstpage}
\pagerange{\pageref{firstpage}--\pageref{lastpage}}
\maketitle

\begin{abstract}
We investigate the impact of local environment on the galaxy stellar mass function (SMF) spanning a wide range of galaxy densities from the field up to dense cores of massive galaxy clusters.
Data are drawn from a sample of eight fields from the Observations of Redshift Evolution in Large-Scale Environments (ORELSE) survey.
Deep photometry allow us to select mass-complete samples of galaxies down to 10$^9$ M$_{\odot}$. 
Taking advantage of $>$4000 secure spectroscopic redshifts from ORELSE and precise photometric redshifts, we construct 3-dimensional density maps between $0.55<z<1.3$ using a Voronoi tessellation approach.
We find that the shape of the SMF depends strongly on local environment exhibited by a smooth, continual increase in the relative numbers of high- to low-mass galaxies towards denser environments.
A straightforward implication is that local environment proportionally increases the efficiency of (a) destroying lower-mass galaxies and/or (b) growth of higher-mass galaxies.
We also find a presence of this environmental dependence in the SMFs of star-forming and quiescent galaxies, although not quite as strongly for the quiescent subsample.
To characterize the connection between the SMF of field galaxies and that of denser environments we devise a simple semi-empirical model.
The model begins with a sample of $\approx$10$^6$ galaxies at $z_{\rm{start}}$=5 with stellar masses distributed according to the field.
Simulated galaxies then evolve down to $z_{\rm{final}}$=0.8 following empirical prescriptions for star-formation, quenching, and galaxy-galaxy merging.
We run the simulation multiple times, testing a variety of scenarios with differing overall amounts of merging.
Our model suggests that a large number of mergers are required to reproduce the SMF in dense environments.
Additionally, a large majority of these mergers would have to occur in intermediate density environments (e.g. galaxy groups).
\end{abstract}

\begin{keywords}
galaxies: evolution -- galaxies: groups: general -- galaxies: clusters: general -- techniques: photometric -- techniques: spectroscopic
\end{keywords}

%%%%%%%%%%%%%%%%%%%%%%%%%%%%%%%%%%%%%%%%%%%%%%%%%%
%%%%%%%%%%%%%%%%% BODY OF PAPER %%%%%%%%%%%%%%%%%%
%%%%%%%%%%%%%%%%%%%%%%%%%%%%%%%%%%%%%%%%%%%%%%%%%%

\section{Introduction}

It has been broadly understood from early observations that galaxies in clusters evolve along different timescales and/or pathways 
relative to those in more isolated environments \citep[e.g.][]{Gunn1972, Oemler1974, Dressler1980}.
A variety of mechanisms either unique to or facilitated in cluster environments have been proposed as a means of contextualizing these observations such as ram-pressure stripping \citep{Gunn1972} and gravitational interactions between galaxies \citep{Richstone1976} and with the cluster potential \citep{Farouki1981}.
In more recent years the Sloan Digital Sky Survey \citep[SDSS: ][]{York2000}, encompassing a cosmologically-significant volume in the local universe, has confirmed the importance of environment in relation to galaxy evolution including the morphology-density relation \citep{Goto2003b}, the colour-density relation \citep{Hogg2004}, and the star-formation rate (SFR) versus density relation \citep{Gomez2003}.
Nevertheless, despite the modern wealth of data from both observations and simulations we still lack a fully coherent picture of how galaxies evolve in dense environments.

One of the most key measurements that can be made for a sample of galaxies is their volume-based number density as a function of stellar mass, known simply as the galaxy stellar mass function (SMF).
All relevant aspects of galaxy evolution that relate to mass-growth and structure formation will naturally be imprinted in some way in this distribution.
Therefore, measuring the SMF provides insight into galaxy evolution as well as being an important diagnostic tool for galaxy simulations.

Over roughly the past decade there has been a multitude of studies dedicated to measuring the galaxy SMF as a function of environment over a broad range of redshifts from $z \sim 1.5$ to the local universe \citep{Bundy2006, Peng2010, Bolzonella2010, Giodini2012, Vulcani2012, Vulcani2013, Calvi2013, vanderBurg2013, Mortlock2015, Hahn2015, Davidzon2016}.
The general picture emerging from many of these studies is that the SMF in denser regions is more heavily weighted towards higher stellar masses, i.e., that the relative number of high- to low-mass galaxies increases with environmental density.
A common theme evoked is that this phenomenon is primarily driven by the changing relative contributions of star-forming and quiescent galaxies to the total galaxy population in different environments.
Recent studies have explored a formalism of quenching involving two pathways, ``mass-quenching'' and ``environment-quenching'', which operate independently of each other \citep{Peng2010, Muzzin2012, Quadri2012}.

As mentioned earlier, several physical mechanisms related to galaxy environment are understood as being relevant for galaxy evolution and are pointed to as drivers of environment-quenching.
However, the physical origin of mass-quenching is less well-understood.
Some suggested mechanisms involve the formation of a stable hot gaseous halo in massive galaxies capable of preventing the accretion of cold gas and effectively quenching a galaxy.
Virial shocks \citep{Birnboim2003} and/or active galactic nuclei \citep{Croton2006} have been proposed as viable energy sources for maintaining such a hot halo.

Combining these two quenching pathways with the observed constancy of the shape of the SMF for star-forming galaxies, \citet{Peng2010} have taken this formalism further to create a model that describes the SMF of quiescent galaxies in terms of two components: mass-quenched galaxies and environment-quenched galaxies.
The mass-quenched component is derived from the requirement of maintaining the shape of the star-forming SMF, which when coupled with the SFR-$M_*$ correlation observed for star-forming galaxies \citep[e.g.][]{Noeske2007} leads to a quenching rate that preferentially affects higher-mass galaxies.
Under the assumption that the environmental quenching efficiency is independent of stellar mass (for which the previously mentioned works provide observational evidence) it follows that the SMF of environment-quenched galaxies will have the same shape as the SMF of star-forming galaxies.
\citet{Peng2010} have shown that this simple, empirically motivated model is able to accurately reproduce the SMF of galaxies in the local universe as measured from the SDSS DR7 \citep{Abazajian2009}.
However, recent works have applied this quenching scheme to measurements of the SMF at $0.5 \lesssim z \lesssim 1$ finding mild to moderate levels of tension \citep{vanderBurg2013, Davidzon2016}.
In the latter of these two studies the authors argue that galaxy-galaxy mergers should also play a significant role in shaping the SMF in high density environments.
%This makes sense intuitively as galaxy-galaxy interactions are expected to occur more frequently in dense environments.
In fact, it has been estimated that the merger rate can be 3-4$\times$ greater in high- versus low-density environments \citep{Lin2010, Kampczyk2013} and exhibits evolution with redshift \citep{LopezSanjuan2013}.
It is important to note that the high-density regime examined in these studies combine both galaxy group and galaxy cluster scales, and are more heavily weighted towards group-like environments.
Indeed galaxy groups are believed to be the environment most conducive to galaxy-galaxy merging due to their moderate velocity dispersions, whereas velocities in clusters may act to suppress merging despite bolstering elevated number densities of galaxies \citep{Lin2010}.

In this paper we present new measurements of the galaxy SMF as a function of local environment from the Observations of Redshift Evolution in Large-Scale Environments survey \citep[ORELSE][]{Lubin2009}.
ORELSE is a wide-field survey dedicated to studying galaxy evolution across the full range of local environments with extensive photometric and spectroscopic observations of multiple well-known galaxy overdensities at $0.6 < z < 1.3$.
In the work presented here we make use of a subset of eight ORELSE fields from the full sample for which all data have been currently reduced.
Nevertheless, this subset covers essentially the entire redshift range of the survey and samples the full range of substructure masses from low-mass groups to rich clusters.

This paper is organized as follows:
In Section \ref{sec:data} we discuss the photometric and spectroscopic data as well as SED-fitting techniques used to derive galaxy properties.
Section \ref{sec:methods} describes the various analysis methodologies employed such as the quantitative definition of local environment and the estimation of mass completeness.
In Section \ref{sec:results} we discuss the results from our measurements and introduce a simple semi-empirical model as an aid in understanding the connection between the SMF in different environments.
Finally, in Section \ref{sec:summaryandconclusions} we summarize our results and consider a future direction for a followup study.
Throughout this paper we adopt a standard $\Lambda$CDM cosmology with $\Omega_{\mathrm{M}} = 0.3$, $\Omega_{\mathrm{\Lambda}} = 0.7$, and $H_0 = 70$ km s$^{-1}$ Mpc$^{-1}$.

\section{Data}
\label{sec:data}

\renewcommand{\thefootnote}{\fnsymbol{footnote}}
In this study we make use of several fields taken from the Observations of Redshift Evolution in Large-Scale Environments survey \citep[ORELSE:][]{Lubin2009}.
ORELSE is a large multi-wavelength photometric and spectroscopic campaign of 16\footnote[2]{Note: the original survey design included 20 separate fields.
However, since then two fields have been merged into the SC 1324 supercluster and another two into the SC 1604 supercluster \citep[see Table 1 of][]{Lubin2009}.
Two other fields (Cl 0943+4804 and Cl 1325+3009) have been removed altogether due to incomplete data acquisition.}
fields, each containing massive large scale structures (LSSs) at $0.6 < z < 1.3$.
The goal of this survey is to characterize galaxies over the full range of environments from sparse fields to the dense cores of rich galaxy clusters.
In the following sections we describe in detail the photometric and spectroscopic data and reductions.
At the present time only eight fields from ORESLE have fully reduced spectroscopy and all available imaging data, all of which are used in the work presented here.
In Table \ref{tab:lss} we list several properties of the LSS fields used in this work.
We note here that most of the LSSs in these fields are not individual galaxy clusters but have been revealed as being composed of multiple galaxy groups and/or clusters \citep[see Section 2 of][]{Rumbaugh2017}.
\renewcommand{\thefootnote}{\arabic{footnote}}

% ~~~~~~~~~~~~~~~~~~~~~~~~~~~~~~~~~~~~~~~~~~~~~~~~~~~~~~~~~~~~~
% ~~~~~~~~~~~~~~~~~~~~~~~~~~~~~ TABLE ~~~~~~~~~~~~~~~~~~~~~~~~~~
% ~~~~~~~~~~~~~~~~~~~~~~~~~~~~~~~~~~~~~~~~~~~~~~~~~~~~~~~~~~~~~

\begin{table}
	\begin{center}
	\caption{Large Scale Structures}
	\label{tab:lss}

	\begin{tabular}{lccccc}

	\hline \\[-5mm]  
	\hline \\[-3mm]  

	Name & R.A.$^a$ & Dec.$^a$ & \zspec$^b$ & N$_{\mathrm{spec}}$$^c$ & log($\Sigma$ M$_{\mathrm{vir}}$)$^d$ \\
	 & J2000 & J2000 & & & [M$_{\odot}$] \\[-0.5mm]

	\hline \\[-2mm]

	RXJ1757  &   269.3319   &   66.5259   &   0.693   &  374   &   14.8  \\
	SC1324   &   201.2143   &   30.1905   &   0.756   &  893   &   15.3  \\
	RCS0224  &   36.14120   &   -0.0394   &   0.772   &  362   &   15.0  \\
	RXJ1716  &   259.2016   &   67.1392   &   0.813   &  372   &   15.3  \\
	RXJ1821  &   275.3845   &   68.4658   &   0.818   &  295   &   15.2  \\
	SC1604   &   241.1409   &   43.3539   &   0.898   &  1150  &   15.4  \\
	SC0910   &   137.5983   &   54.3419   &   1.110   &  430   &   15.0  \\
	SC0849   &   132.2333   &   44.8711   &   1.261   &  344   &   15.0  \\[0.5mm]
	\hline

	\end{tabular}
	\end{center}

	$^a$ Central coordinates of the main LSS in each field. \\
	$^b$ Mean redshift of the LSS \\
	$^c$ Total number of secure spectroscopic redshifts at $0.55 < z < 1.3$ \\
	$^d$ Note that for many of these LSSs we detect multiple galaxy groups and/or clusters \citep[see][]{Rumbaugh2017}.
	In these cases we show the sum of the virial masses of all substructures.
\end{table}

% ~~~~~~~~~~~~~~~~~~~~~~~~~~~~~~~~~~~~~~~~~~~~~~~~~~~~~~~~~~~~~
% ~~~~~~~~~~~~~~~~~~~~~~~~~~~~~ TABLE ~~~~~~~~~~~~~~~~~~~~~~~~~~
% ~~~~~~~~~~~~~~~~~~~~~~~~~~~~~~~~~~~~~~~~~~~~~~~~~~~~~~~~~~~~~

\begin{table*}
	\begin{center}
	\caption{Photometry}
	\label{tab:photometry}

	{\vskip 1mm}
	\begin{tabular}{c @{\hskip 15mm} c}

		\begin{tabular}{llll}

		\hline \\[-3.3mm]  

		Filter & Telescope & Instrument & Depth$^a$ \\[0mm]

		\hline \\[-3mm]  
		SC1604 \\[-1mm]
		\hline \\[-5mm]  
		\hline \\[-2mm]

		$B$    &   Subaru   &   Suprime-Cam   &   26.6   \\
		$V$    &   Subaru   &   Suprime-Cam   &   26.1   \\
		$R_C$  &   Subaru   &   Suprime-Cam   &   26.0   \\
		$I_C$  &   Subaru   &   Suprime-Cam   &   25.1   \\
		$Z_+$  &   Subaru   &   Suprime-Cam   &   24.6   \\
		$r'$   &   Palomar   &   LFC   &   24.2   \\
		$i'$   &   Palomar   &   LFC   &   23.6   \\
		$z'$   &   Palomar   &   LFC   &   23.1   \\
		$J$   &   UKIRT   &   WFCAM   &   22.1   \\
		$K$   &   UKIRT   &   WFCAM   &   21.9   \\
		$[3.6]$  &   {\it Spitzer}  &   IRAC   &   24.7   \\
		$[4.5]$  &   {\it Spitzer}  &   IRAC   &   24.3   \\
		$[5.8]$  &   {\it Spitzer}  &   IRAC   &   22.7   \\
		$[8.0]$  &   {\it Spitzer}  &   IRAC   &   22.6   \\[1mm]

		\hline \\[-3mm]
		RXJ1716 \\[-1mm]
		\hline \\[-5mm]  
		\hline \\[-2mm]

		$B$    &   Subaru   &   Suprime-Cam   &   25.9   \\
		$V$    &   Subaru   &   Suprime-Cam   &   26.6   \\
		$R_C$  &   Subaru   &   Suprime-Cam   &   26.2   \\
		$I_+$  &   Subaru   &   Suprime-Cam   &   25.4   \\
		$Z_+$  &   Subaru   &   Suprime-Cam   &   24.7   \\
		$J$    &   CFHT   &   WIRCam   &   21.3   \\
		$K_s$  &   CFHT   &   WIRCam   &   21.7   \\
		$[3.6]$  &   {\it Spitzer}  &   IRAC   &   24.6   \\
		$[4.5]$  &   {\it Spitzer}  &   IRAC   &   24.1   \\
		$[5.8]$  &   {\it Spitzer}  &   IRAC   &   22.4   \\
		$[8.0]$  &   {\it Spitzer}  &   IRAC   &   22.3   \\[1mm]

		% aka nep200
		\hline \\[-3mm]
		RXJ1757 \\[-1mm]
		\hline \\[-5mm]  
		\hline \\[-2mm]

		$B$    &   Subaru   &   Suprime-Cam   &   26.4   \\
		$V$    &   Subaru   &   Suprime-Cam   &   25.9   \\
		$R_C$  &   Subaru   &   Suprime-Cam   &   26.7   \\
		$Z_+$  &   Subaru   &   Suprime-Cam   &   25.6   \\
		$r'$   &   Palomar   &   LFC   &   25.1   \\
		$i'$   &   Palomar   &   LFC   &   24.8   \\
		$z'$   &   Palomar   &   LFC   &   22.9   \\
		$Y$    &   Subaru   &   Suprime-Cam   &   22.7   \\
		$J$     &   CFHT   &   WIRCam   &   21.0   \\
		$K_s$   &   CFHT   &   WIRCam   &   21.8   \\
		$[3.6]$  &   {\it Spitzer}  &   IRAC   &  23.9    \\
		$[4.5]$  &   {\it Spitzer}  &   IRAC   &  23.8    \\[1mm]

		\hline \\[-3mm]
		RCS0224 \\[-1mm]
		\hline \\[-5mm]  
		\hline \\[-2mm]

		$B$    &   Subaru   &   Suprime-Cam   &   26.2   \\
		$V$    &   Subaru   &   Suprime-Cam   &   26.0   \\
		$R_+$    &   Subaru   &   Suprime-Cam   &   25.9   \\
		$I_+$    &   Subaru   &   Suprime-Cam   &   25.5   \\
		$Z_+$    &   Subaru   &   Suprime-Cam   &   24.9   \\
		$J$   &   UKIRT   &   WFCAM   &   21.2   \\
		$K$   &   UKIRT   &   WFCAM   &   21.4   \\
		$[3.6]$  &   {\it Spitzer}  &   IRAC   &   24.0   \\
		$[4.5]$  &   {\it Spitzer}  &   IRAC   &   23.6   \\

		\end{tabular}

		&

		\begin{tabular}{llll}

		\hline \\[-3mm]  

		Filter & Telescope & Instrument & Depth$^a$ \\[0mm]

		\hline \\[-3mm]  
		SC0849 \\[-1mm]
		\hline \\[-5mm]  
		\hline \\[-2mm]

		$B$    &   Subaru   &   Suprime-Cam   &   26.4   \\
		$V$    &   Subaru   &   Suprime-Cam   &   26.5   \\
		$R_C$  &   Subaru   &   Suprime-Cam   &   26.2   \\
		$I_+$  &   Subaru   &   Suprime-Cam   &   25.5   \\
		$Z_+$  &   Subaru   &   Suprime-Cam   &   25.1   \\
		$Z_R$  &   Subaru   &   Suprime-Cam   &   23.5   \\
		$r'$   &   Palomar   &   LFC   &   24.7   \\
		$i'$   &   Palomar   &   LFC   &   24.4   \\
		$z'$   &   Palomar   &   LFC   &   23.3   \\
		$NB711$  &   Subaru   &   Suprime-Cam   &   23.7   \\
		$NB816$  &   Subaru   &   Suprime-Cam   &   25.9   \\
		$J$   &   UKIRT   &   WFCAM   &   21.8   \\
		$K$   &   UKIRT   &   WFCAM   &   21.6   \\
		$[3.6]$  &   {\it Spitzer}  &   IRAC   &   24.8   \\
		$[4.5]$  &   {\it Spitzer}  &   IRAC   &   24.3   \\[1mm]

		\hline \\[-3mm]
		SC1324 \\[-1mm]
		\hline \\[-5mm]  
		\hline \\[-2mm]

		$B$    &   Subaru   &   Suprime-Cam   &   26.6   \\
		$V$    &   Subaru   &   Suprime-Cam   &   25.7   \\
		$R_C$  &   Subaru   &   Suprime-Cam   &   25.7   \\
		$I_+$  &   Subaru   &   Suprime-Cam   &   25.2   \\
		$Z_+$  &   Subaru   &   Suprime-Cam   &   22.6   \\
		$r'$   &   Palomar   &   LFC   &   24.9   \\
		$i'$   &   Palomar   &   LFC   &   24.3   \\
		$z'$   &   Palomar   &   LFC   &   22.6   \\
		$J$   &   UKIRT   &   WFCAM   &   22.4   \\
		$K$   &   UKIRT   &   WFCAM   &   21.7   \\
		$[3.6]$  &   {\it Spitzer}  &   IRAC   &   23.9   \\
		$[4.5]$  &   {\it Spitzer}  &   IRAC   &   23.8   \\[1mm]

		% aka nep5281
		\hline \\[-3mm]
		RXJ1821 \\[-1mm]
		\hline \\[-5mm]  
		\hline \\[-2mm]

		$B$    &   Subaru   &   Suprime-Cam   &   26.0   \\
		$V$    &   Subaru   &   Suprime-Cam   &   26.0   \\
		$r'$   &   Palomar   &   LFC   &   24.4   \\
		$i'$   &   Palomar   &   LFC   &   24.3   \\
		$z'$   &   Palomar   &   LFC   &   23.3   \\
		$Y$    &   Subaru   &   Suprime-Cam   &   23.4   \\
		$J$     &   CFHT   &   WIRCam   &   21.4   \\
		$K_s$   &   CFHT   &   WIRCam   &   21.7   \\
		$[3.6]$  &   {\it Spitzer}  &   IRAC   &   23.9   \\
		$[4.5]$  &   {\it Spitzer}  &   IRAC   &   23.8   \\[1mm]

		\hline \\[-3mm]
		SC0910 \\[-1mm]
		\hline \\[-5mm]  
		\hline \\[-2mm]

		$B$    &   Subaru   &   Suprime-Cam   &   24.4   \\
		$V$    &   Subaru   &   Suprime-Cam   &   25.6   \\
		$R_C$  &   Subaru   &   Suprime-Cam   &   26.4   \\
		$I_+$  &   Subaru   &   Suprime-Cam   &   25.8   \\
		$Z_+$  &   Subaru   &   Suprime-Cam   &   24.8   \\
		$J$   &   UKIRT   &   WFCAM   &   22.1   \\
		$K$   &   UKIRT   &   WFCAM   &   21.7   \\
		$[3.6]$  &   {\it Spitzer}  &   IRAC   &   23.2   \\
		$[4.5]$  &   {\it Spitzer}  &   IRAC   &   23.2   \\[1mm]

		\end{tabular}

	\end{tabular}

	\end{center}

	$^a$ 80\% completeness limits derived from the recovery rate of artificial sources inserted at empty sky regions.

\end{table*}

\subsection{Photometry}

Each LSS field comprises a wealth of photometric data spanning optical to mid-infrared wavelengths.
These imaging data were compiled from both proposed observing campaigns part of ORELSE as well as archival data from Suprime-Cam \citep{Miyazaki2002} on Subaru, the Large Format Camera \citep[LFC:][]{Simcoe2000} on the Palomar 200-inch Hale telescope, the Wide-field InfraRed Camera \citep[WIRCam:][]{Puget2004} on the Canada France Hawaii Telescope (CFHT), the Wide Field Camera \citep[WFCAM:][]{Casali2007} on the United Kingdom InfraRed Telescope (UKIRT), and the InfraRed Array Camera \citep[IRAC:][]{Fazio2004} on the {\it Spitzer Space Telescope}.
Table \ref{tab:photometry} lists the available photometry for each field, which facilities and instruments were used, as well as depth estimates.

Optical and Y-band imaging from Subaru Suprime-Cam was reduced using the SDFRED pipeline\footnote{Note that there are two versions of this software: SDFRED1 and SDFRED2 for data taken before and after July 2008 respectively. See \url{http://subarutelescope.org/Observing/Instruments/SCam/sdfred/index.html.en}} \citep{Ouchi2004} as well as several Terapix\footnote{\url{http://terapix.iap.fr}} software packages.
Basic image processing is first done with SDFRED which we use to perform (1) bias level subtraction based on designated overscan regions on the Suprime-Cam CCDs, (2) flat field correction,
(3) optics and atmospheric distortion correction, and (4) equalization of the point spread function (PSF).
Flat fields were constructed directly from the science images by masking out objects to create ``super sky flats''.
This is possible due to the nature of the observations which include many individual dithered images and because we do not observe large extended objects.
Next we perform astrometric alignment, photometric scaling, and image stacking using the Source Extractor \citep[SExtractor:][]{Bertin1996}, Software for Calibrating AstroMetry and Photometry \cite[SCAMP:][]{Bertin2006}, and SWarp \citep{Bertin2002} software packages to create fully-reduced mosaics and weight maps.
Photometric calibration is performed from same-night observations of standard star fields from the \citet{Landolt1992} catalogs. 
Standard fields are reduced identically as described above.

Optical imaging from the Large Format Camera (LFC) mounted on the 200-inch Hale Telescope at Palomar Observatory is reduced using a suite of scripts\footnote{\url{http://www.ifa.hawaii.edu/~rgal/science/lfcred/lfc_red.html}} written in Image Reduction and Analysis Facility (IRAF).
These scripts perform standard procedures for image processing (e.g. bias subtraction, flat field correction) as well as a few routines written specifically for handling LFC data.
We refer the reader to \citet{Gal2005} for a more detailed discussion of the processing of LFC images.

Near-infrared imaging in the $J$ and $K$/$K_s$ bands were taken using UKIRT/WFCAM and CFHT/WIRCam.
Both facilities implement automated data reduction pipelines that output fully-reduced mosaics and weight maps.
We photometrically calibrate these mosaics using bright ($<$15 mag), non-saturated objects with existing photometry from 2MASS in each field.
It is important to note that the 2MASS $K_s$ filter is significantly different than the WFCAM $K$ filter.
In order to account for this we make use of infrared spectral observations of stars from the Infrared Telescope Facility (IRTF) spectral library \citep{Rayner2009} to calculate a transformation between these two passbands $(K_{\mathrm{WFCAM}} - K_{s, \mathrm{2MASS}})$ based on their $(J - K_s)$ colours.
%{\bf \color{red}
%It is important to note that the 2MASS $K_s$ filter is significantly different than the WFCAM $K$ filter.
%In order to account for this we make use of infrared spectral observations of stars from the Infrared Telescope Facility (IRTF) spectral library (Rayner et al. 2009)\footnote{\url{http://irtfweb.ifa.hawaii.edu/~spex/IRTF_Spectral_Library/}} to calculate a transformation between these two %passbands $(K_{\mathrm{WFCAM}} - K_{s, \mathrm{2MASS}})$ based on their $(J - K_s)$ colours.
%}

Once all ground-based optical and near-infrared images (B-K bands) have been assembled for a given field we register them to a common pixel grid (0.2 \arcsec /pixel).
We account for small astrometric distortions, such as offsets and rotations, among the final mosaics using the SCAMP software.
Due to differences in the imaging for each LSS field we carefully choose detection images on an ad-hoc basis.
In some cases we use an inverse-variance weighted stack of two or more images for source detection.
For each field we require that the detection image be predominantly redward of the 4000\AA\ break at the \z\ of the LSS contained in it, have sufficiently good seeing ($<1$\arcsec ), and be sufficiently deep to detect low-mass galaxies (completeness limits will be discussed in section \ref{sec:masslimits}).
We make an exception for the SC1604 field for which only the $R_C$ image covered the full extent of the spectroscopic footprint.
Table \ref{tab:masslim} lists images used for source detection as well as their seeing.

Point spread functions are created for each image by stacking a selection of bright, non-saturated point sources.
We then use the Richardson \& Lucy algorithm in {\tt scikit-image} \citep{vanderWalt2014arxiv} to construct convolution kernels which are used to smooth all optical-NIR images to the largest PSF.
Photometry is extracted from PSF-matched images in fixed circular apertures by running SExtractor in dual-image mode using the aforementioned detection images.
Aperture diameters are chosen to be 1.3$\times$ the full-width half-maximum (FWHM) of the largest PSF, where the signal-to-noise ratio (S/N) of the extracted flux is near its maximum \citep[see Section 4.1 of][]{Whitaker2011}.
Fluxes are corrected for attenuation from dust in the Milky Way based on the reddening maps derived by \citet{Schlafly2011} as provided by the NASA/IPAC Infrared Science Archive\footnote{\url{http://irsa.ipac.caltech.edu/applications/DUST/}} (IRSA).
We define an aperture correction as the ratio of an object's total flux to its aperture flux in the PSF-matched detection image, where we use SExtractor's {\tt AUTO} flux as the ``total'' flux.

All of our cluster fields have non-cryogenic {\it Spitzer}/IRAC imaging ($[3.6]$ and $[4.5]\mu$m).
However, two fields (RXJ1716 and SC1604) were observed during the cryogenic mission and thus have $[5.8]$ and $[8.0]\mu$m imaging.
We process the individual corrected basic calibrated data (cBCD) images using the {\tt MOPEX} software package \citep{Makovoz2006} along with several custom scripts.
We first preprocess the cBCD frames using a custom IDL code (J. Surace, private communication) which performs and improved correction for column pulldown and muxbleed artifacts.
The images are next background-matched using the {\tt overlap.pl} routine from {\tt MOPEX} prior to mosaicing with the {\tt mosaic.pl} routine.
All {\it Spitzer} imaging is flux-calibrated and provided in units of MJy/sr.

% ~~~~~~~~~~~~~~~~~~~~~~~~~~~~~~~~~~~~~~~~~~~~~~~~~~~~~~~~~~~~~
% ~~~~~~~~~~~~~~~~~~~~~~~~~~~~~ TABLE ~~~~~~~~~~~~~~~~~~~~~~~~~
% ~~~~~~~~~~~~~~~~~~~~~~~~~~~~~~~~~~~~~~~~~~~~~~~~~~~~~~~~~~~~~

\begin{table*}
	\begin{center}
	\caption{Image Properties \& Completeness Limits}
	\label{tab:masslim}

	\begin{tabular}{lccccccc}

	\hline \\[-5mm]  
	\hline \\[-3mm]  

	Name & Detection & FWHM & Area$^a$ & M$_{\mathrm{*, lim}}$$^b$ & M$_{\mathrm{*, ssp}}$$^c$ & $\sigma_{\Delta z / (1 + z_{\mathrm{spec}})}$ & f$_{\mathrm{outlier}}$ \\
	 & Image & [ \arcsec\ ] & [ arcmin$^2$ ] & [M$_{\odot}$] & [M$_{\odot}$] & [\%] & [\%] \\[0.5mm]

	\hline \\[-2mm]

	RXJ1757   &  $r', i'$         &  0.79  &  203  &   9.21   &   9.72    &   3.0   &   8.7  \\
	SC1324    &  $i'$             &  0.97  &  465  &   9.46   &   10.02   &   3.2   &   9.2  \\
	RCS0224   &  $I_+$            &  0.67  &  172  &   8.73   &   9.74    &   2.8   &   3.1  \\
	RXJ1716   &  $R_C, I_+, Z_+$  &  0.69  &  194  &   9.05   &   9.70    &   2.2   &   5.4  \\
	RXJ1821   &  $Y$              &  0.83  &  179  &   9.73   &   10.01   &   3.1   &   4.8  \\
	SC1604    &  $R_C$            &  1.00  &  270  &   9.28   &   10.05   &   2.9   &   9.5  \\
	SC0910    &  $R_C, I_+, Z_+$  &  0.65  &  185  &   9.52   &   10.13   &   3.2   &   8.1  \\
	SC0849    &  $Z_+$            &  0.77  &  147  &   9.46   &   10.19   &   2.7   &   6.0  \\[0mm]
	\hline

	\end{tabular}
	\end{center}

	$^a$ Total projected area covered by spectroscopic observations \\
	$^b$ 80\% mass completeness limit from artificial sources \\
	%, see section \ref{sec:masslimits} for details \\
	%exponentially declining SFH tau=1Gyr
	%zform = 5
	%no dust
	$^c$ Mass limit from a dust-free SSP with $z_{\mathrm{form}} = 5$\\
\end{table*}

Due to the significantly larger PSFs of {\it Spitzer}/IRAC images ($\sim 2$\arcsec ) we take an alternate approach for measuring photometry.
We use {\tt T-PHOT} \citep{Merlin2015}, a software package designed to extract photometry in crowded images where blending from neighbours is significant.
In brief the methodology of {\tt T-PHOT} is as follows: first, the positions and morphologies of objects are obtained for use as priors based on a segmentation map (produced by SExtractor) of a higher-resolution image, in our case the detection image mentioned above.
Cutouts are taken from this higher-resolution image which are then used to create low-resolution models of each object by smoothing with a provided convolution kernel.
These models are then simultaneously fit to the IRAC image until optimal scale factors, assessed through a global $\chi^2$ minimization, for each object are obtained.
We run {\tt T-PHOT} in ``{\it cells-on-objects}'' mode where when fitting a model for a given object only neighbours within a 1$\times$1\arcmin\ box centred on the object are considered in the fitting.
After this initial sequence, {\tt T-PHOT} is then rerun in a second pass in which registered kernels generated during the first pass are utilized to account for mild astrometric differences between the input images.
Output fluxes are the total model flux from the best fit.
Therefore, in order to make these fluxes consistent with the fixed-aperture photometry of the optical-NIR imaging we scale the IRAC flux of a given object by the ratio of the aperture to total flux from the PSF-matched detection image.
We use the SExtractor {\tt AUTO} aperture flux for the ``total'' flux.

\subsection{Spectroscopy}
\label{sec:spectroscopy}

The bulk of the spectroscopic data used in this study were obtained as part of a massive 300 hour \ion{Keck}{2}/DEep Imaging Multi-Object Spectrograph 
(DEIMOS; \citealt{Faber2003}) campaign targeting all fields of the ORELSE survey listed in Table 1 of \citet{Lubin2009}. In this section we briefly describe the targeting, acquisition, and 
reduction of those data pertaining to the eight fields presented in this study. A total of 67 slitmasks were observed over the eight fields, with the number
of slitmasks per field ranging from four (RCS0224) to 18 (SC1604). Slitmasks were observed with the 1200 line mm$^{-1}$ grating with 1$\arcsec$ slits resulting
in a plate scale of 0.33\AA\ pix$^{-1}$, an $R\sim5000$ ($\lambda$/$\theta_{\mathrm{FWHM}}$, where $\theta_{\mathrm{FWHM}}$ is the full-width half-maximum resolution), and 
a wavelength coverage of $\Delta\lambda$$\sim$2600. Central wavelengths ranged from $7200\rm{\AA} \le \lambda_{c} \le 8700\rm{\AA}$ depending on the 
redshift of the field and exposure times varied from 3600-10800s, scaled to roughly achieve an identical distribution in continuum S/N across all masks
independent of conditions and the median faintness of the target population. For more details
on the range of conditions, setups, and exposure times for each individual field see \citet{Rumbaugh2017}. In addition to the DEIMOS campaign,
a small fraction ($\la3$\%) of the spectroscopic redshifts used in this study were drawn from a variety of other studies	% Adam, I took a guess here, change it if it's wrong. 
\citep{Oke1998, Gal2004, Tanaka2008, Mei2012} which utilized a variety of different telescopes and instruments. For the spectroscopic redshifts 
drawn from these studies, we imposed, to the best of our ability, similarly stringent criteria as those imposed on the DEIMOS data (see below) in order to 
consider a redshift secure.

Targets for DEIMOS slitmasks were selected primarily based on the observed-frame colour-colour cuts presented in \citet{Lubin2009}. Objects which satisfied 
the observed-frame $r^{\prime}-i^{\prime}$ and $i^{\prime}-z^{\prime}$ colour range for the redshift of the targeted LSS (see Table 2 of \citealt{Lubin2009}), a colour
range designed to select galaxies on or near the red sequence at that redshift, were given the highest targeting priority (priority 1). Objects of progressively 
bluer colours were assigned progressively lower targeting priorities.
It is important to note that while redder objects were heavily favored in our selection scheme, because of the strictness of our cuts 
and the relative rarity of objects at these colours, the majority of spectroscopic targets had colours bluer than the cuts used to define priority 1 objects. 
The average fraction of priority 1 objects per mask varied considerably from field to field, from 10.0\% in RX J0910 to 45.4\% in RCS0224, with more well-sampled fields 
generally having a smaller average priority 1 fraction. This colour-colour selection scheme was augmented by additional prioritization designed 
to select targets of particular interest that included multiwavelength objects (i.e., radio or X-ray) and known or possible member galaxies
targeted by previous surveys. Targeted objects were generally restricted to $i^{\prime}<24.5$, though this was not a hard limit. 
While this colour-colour preselection was designed to target members of the main LSS in each field, a significant number of priority 2+ objects were also observed, helping to sample the full colour space of galaxies \citep[see][for a full description]{Lubin2009}.
Indeed, as was shown by \citet{Shen2017arxiv}, the ORELSE spectroscopic sample is broadly representative of the underlying galaxy population over a wide range of stellar masses and colours.

Data were reduced using a version of the Deep Evolutionary Extragalactic Probe 2 (DEEP2; \citealt{Davis2003, Newman2013}) \texttt{spec2d} pipeline modified
to improve the precision of the response correction, perform absolute spectrophotometic flux calibration, and improve the handling of the joining
of the blue and red ends of the spectra over the $\sim5$\AA\ gap that separates the two CCD arrays.  All objects, those targeted and those which 
serendipitously fell in a slit, were visually inspected and assigned a spectroscopic redshift (hereafter \zspec ) and a redshift quality code 
($Q$) in the \texttt{zspec} environment following the DEEP2 scheme (see \citealt{Gal2008, Newman2013}). The process for finding and extracting serendipitous 
detections and assigning photometric counterparts is detailed in \citet{Lemaux2009}. Spectroscopic redshifts were assigned a quality code of 
$Q$=-1 (stars) or $Q=3,4$ (galaxies) were considered secure. To be identified securely as a star we required either the presence of multiple significant
narrow photospheric absorption features (typically H$\alpha$ and the \ion{Ca}{2} triplet) or obvious broad continuum features indicative of a late-type star
(primarily TiO). An extragalactic spectroscopic redshifts was considered secure only in the presence of two or more emission or absorption features, 
where $Q=3$ indicates that one or more of the features was slightly questionable in S/N or for some other reason. The two components of the 
$\lambda$3726\AA, $\lambda$3729\AA\ [\ion{O}{2}] feature were considered sufficient to assign a spectrum $Q=4$ when unblended as long as 
both components were significantly detected. A quality code of $Q=3$ was assigned when the two components moderately blended by velocity effects
in the absence of additional corroborating features. It has been shown that adopting this scheme and these quality code cuts results in 
extragalactic redshift measurements that are reliable at the $\ga$99\% level \citep{Newman2013}. For more details on the meaning of the quality 
codes and the likelihood for a redshift of a given quality code to be accurate see \citet{Newman2013}.

\subsection{Spectral Energy Distribution Fitting}
\label{sec:sedfitting}

We perform spectral energy distribution (SED) fitting on the observed optical through mid-IR photometry in order to derive photometric redshifts as well as physical properties (e.g. stellar mass, A$_{\mathrm{V}}$).
In estimating photometric redshifts we employ the code Easy and Accurate Redshifts from Yale \citep[{\tt EAZY}:][]{Brammer2008}.
For this we use fixed-aperture fluxes measured from PSF-matched images.
Briefly, {\tt EAZY} utilizes a set of 6 basis template SEDs derived from a non-negative matrix factorization decomposition of the Projet d'Etude des GAlaxies par Synth\'ese Evolutive template library \citep[P\'EGASE:][]{Fioc1997}, as well as one additional template representing an old stellar population from the \citet{Maraston2005} library (this last template was added in order to help accommodate evolved galaxies at $z<1$ which are not well-fit by the 6 basis templates).
{\tt EAZY} performs $\chi^2$ minimization at points along a user-defined redshift grid using linear combinations of this default template set.
A probability density function (PDF) is then calculated from the minimized $\chi^2$ values: $\mathrm{P}(z) \propto e^{-\chi^2 / 2}$.
Finally, this PDF is modulated by a magnitude prior ($R$-band for this work) which is designed to mimic the intrinsic redshift distribution for galaxies of given apparent magnitude.
{\tt EAZY} is capable of reporting multiple different types of photometric redshifts which are derived from different manipulations of the final redshift PDF.
Throughout this paper we adopt {\tt EAZY}'s ``\zpeak'' for the photometric redshifts of galaxies.
\zpeak\ is obtained by marginalizing over the final PDF.
Additionally, if an object has multiple peaks in its PDF, EAZY will only marginalize over the peak with the largest integrated probability.

% ~~~~~~~~~~~~~~~~~~~~~~~~~~~~~~~~~~~~~~~~~~~~~~~~~~~~~~~~~~~~~
% ~~~~~~~~~~~~~~~~~~~~~~~~~~~~~ FIGURE ~~~~~~~~~~~~~~~~~~~~~~~~~
% ~~~~~~~~~~~~~~~~~~~~~~~~~~~~~~~~~~~~~~~~~~~~~~~~~~~~~~~~~~~~~
\begin{figure*}
	\includegraphics[width=2\columnwidth]{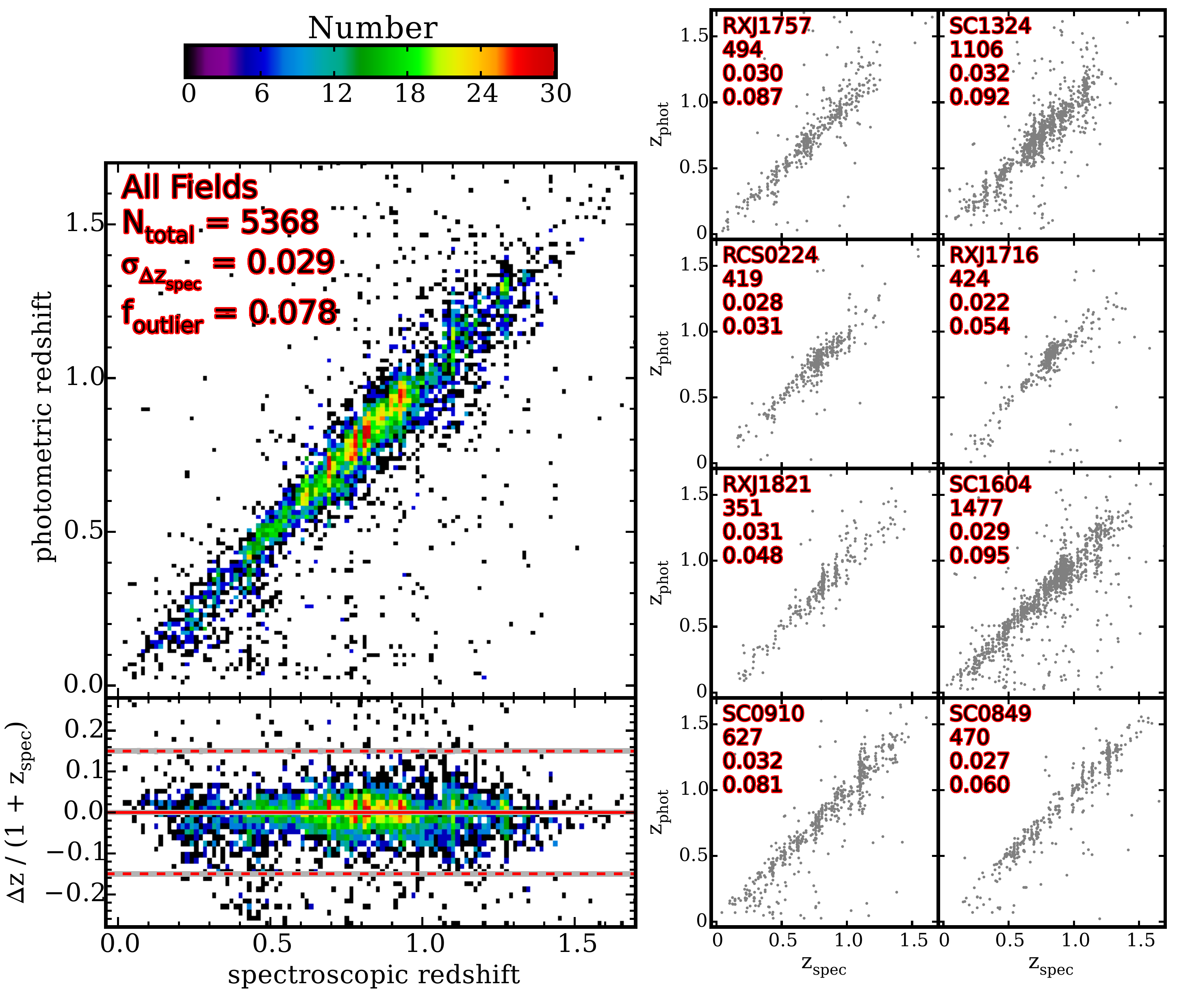}
    \caption{
	A plot of spectroscopic versus photometric redshifts for the LSS fields of ORELSE used in this work.
	Indicated in the top-left corner are the total number of galaxies, the $1\sigma$ scatter derived from fitting a Gaussian to the residual $\Delta z / (1 + z_{\mathrm{spec}})$, and the catastrophic outlier fraction ($|\Delta z | / (1 + z_{\mathrm{spec}}) \geq 0.15$).
	Subpanels on the right show the same distribution and statistics for each field individually.
	}
	\label{fig:zphotzspec}
\end{figure*}

% ~~~~~~~~~~~~~~~~~~~~~~~~~~~~~~~~~~~~~~~~~~~~~~~~~~~~~~~~~~~~~
% ~~~~~~~~~~~~~~~~~~~~~~~~~~~~~ FIGURE ~~~~~~~~~~~~~~~~~~~~~~~~~
% ~~~~~~~~~~~~~~~~~~~~~~~~~~~~~~~~~~~~~~~~~~~~~~~~~~~~~~~~~~~~~
\begin{figure*}
	\includegraphics[width=2\columnwidth]{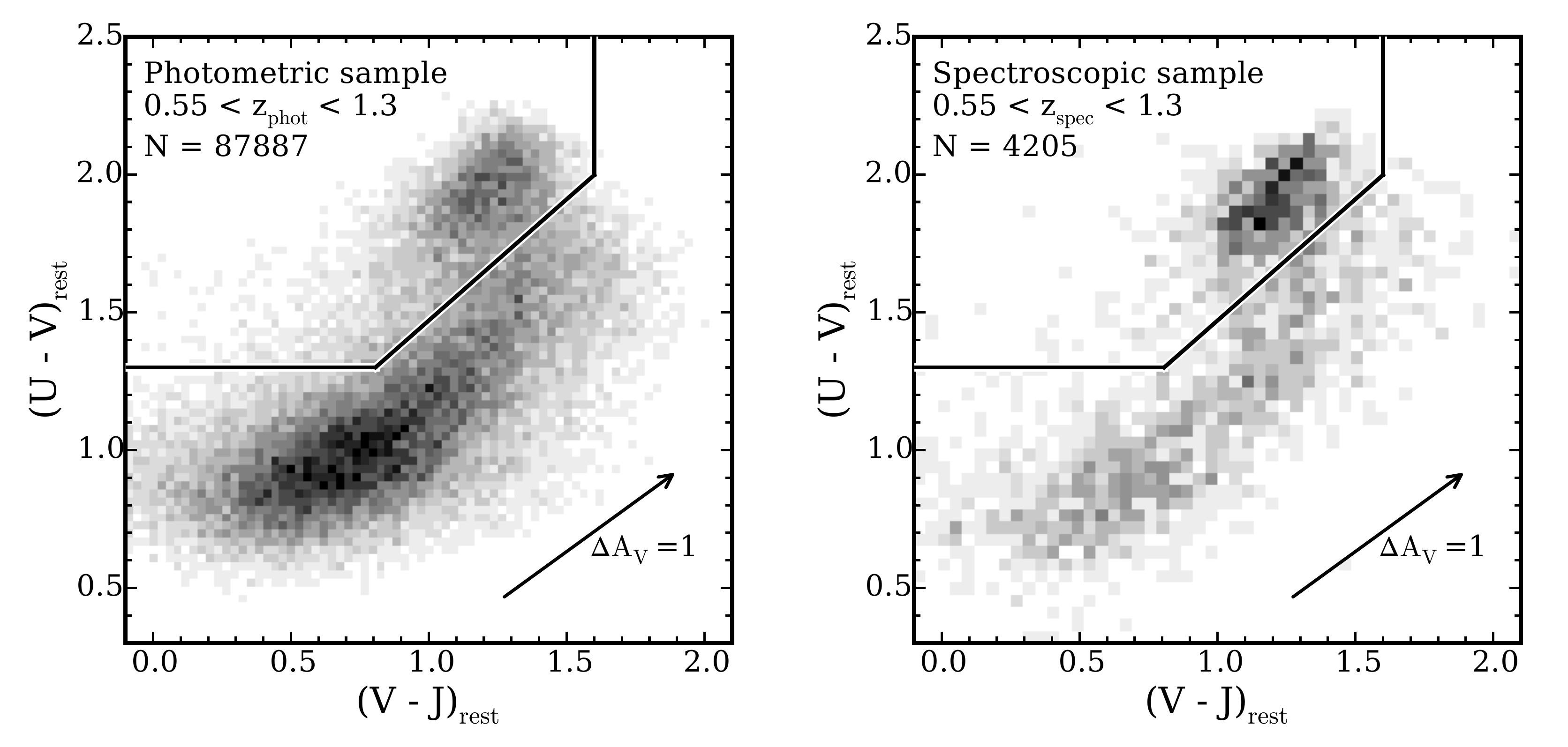}
	\caption{
	Rest-frame $U-V$ versus $V-J$ colour diagram used to classify galaxies as star-forming or quiescent.
	Only galaxies at $0.55 < z < 1.3$ above the estimated stellar mass completeness limits with good use-flags are shown.
	The left and right panels show the photometric and spectroscopic samples respectively, with total numbers of galaxies indicated in the top-left corner.
	%Spectroscopic targeting was designed to prioritize galaxies that are probable members of the LSSs in each field \citep[see][]{Lubin2009} which explains the elevated quiescent fraction relative to the photometric sample.
	Spectroscopic targeting was designed to prioritize galaxies that are probable members of the LSSs in each field \citep[see][]{Lubin2009} which explains the elevated quiescent fraction relative to the photometric sample.
	%The main advantage of this diagram over using a single rest-frame colour is that it is robust against  misclassifying star-forming galaxies reddened by dust as quiescent, illustrated by the arrow which shows the reddening vector caused by $\Delta \mathrm{A_V} = 1$ following the \citet{Calzetti2000} extinction law.
	The main advantage of this diagram over using a single rest-frame colour is that it is robust against  misclassifying star-forming galaxies reddened by dust as quiescent, illustrated by the arrow which shows the reddening vector caused by $\Delta \mathrm{A_V} = 1$ following the \citet{Calzetti2000} extinction law.
	}
	\label{fig:uvj}
\end{figure*}

To assess the precision and accuracy of these photometric redshifts we compare to our spectroscopic redshift measurements.
In Figure \ref{fig:zphotzspec} we plot \zphot\ vs. \zspec\ for the sample of galaxies with high-quality spectroscopic redshifts.
For each field we fit a Gaussian to the distribution of the residual $(z_{\mathrm{phot}} - z_{\mathrm{spec}}) / (1 + z_{\mathrm{spec}})$ for all objects and take the best-fit $\sigma_{\Delta z / (1 + z_{\mathrm{spec}})}$ as the \zphot\ uncertainty.
Values of $\sigma_{\Delta z / (1 + z_{\mathrm{spec}})}$, ranging between 2.2-3.2\%, are shown in Figure \ref{fig:zphotzspec} and Table \ref{tab:masslim}.

We next utilize the code Fitting and Assessment of Synthetic Templates \citep[{\tt FAST}:][]{Kriek2009} for deriving stellar masses and other physical properties.
Here we use aperture-corrected fluxes
and fix galaxies to their derived photometric redshift (or secure \zspec\ when available).
In brief, FAST creates a multi-dimensional cube of model fluxes from a provided stellar population synthesis library (SPS).
Each object in the photometric catalog is fit by every model in this cube and the minimum $\chi^2$ for each is recorded.
The model with the lowest minimum $\chi^2$ is adopted as the best-fit.
In this work we use the SPS library of \citet{Bruzual2003} (BC03) assuming a \citet{Chabrier2003} stellar initial mass function (IMF) and solar metallicity.
For dust extinction we adopt the \citet{Calzetti2000} attenuation curve.
We adopt delayed exponentially declining star-formation histories ($\mathrm{SFH} \propto t \times e^{-t / \tau}$), allowing log($\tau/\mathrm{yr}$) to range between 7-10 in steps of 0.2, log($age/\mathrm{yr}$) to range between 7.5-10.1 in steps of 0.1, and $\mathrm{A_V}$ to range between 0-4 in steps of 0.1.
These assumptions are broadly consistent with other contemporary multi-wavelength extragalactic surveys.

To aid in the selection of real galaxies within our catalogs we define a ``use'' flag for objects in similar fashion to recent multi-wavelength photometric surveys \citep{Whitaker2011, Muzzin2013a, Skelton2014, Straatman2016}.
This use-flag is designed to reject objects that are either poorly detected ($S/N < 3$), saturated, have catastrophic SED fits, or are likely foreground stars.
Catastrophic SED fits are defined as objects with reduced $\chi^2_{\mathrm{galaxy}} > 10$ from fitting with {\tt EAZY}.
This threshold was chosen arbitrarily based on visual inspection of individual SED fits as well as distributions of derived rest-frame colours of objects.
For identifying foreground stars we employ a separate set of criteria.
First, we perform another round of SED fitting with {\tt EAZY} using the stellar template library of \citet{Pickles1998}.
These stellar templates are fit in single-template mode with no redshifting applied.
An object is flagged as a star in our catalogs if it:

\begin{enumerate}
\item[1)] is detected at $S/N > 3$ \\[-4mm]
\item[2)] has a diameter that is $<1.3 \times$ the FWHM of the PSF \\[-4mm]
\item[3)] has a major- to minor-axis ratio $<1.1$ \\[-4mm]
\item[4)] has a reduced $\chi^2_{\mathrm{stellar}} < \chi^2_{\mathrm{galaxy}}$ \\[-4mm]
\item[5)] is not flagged as a galaxy based on its observed spectrum
\end{enumerate}

Furthermore, any object which happens to not meet all of these criteria but has been flagged as a star based on the spectroscopic observations discussed in Section \ref{sec:spectroscopy} will automatically be rejected as a star.

\section{Methods}
\label{sec:methods}

\subsection{Completeness Limits}
\label{sec:masslimits}

In this section we describe how photometric and stellar mass completeness limits are derived.
For each image we estimate the photometric depth by performing simulations.
Artificial sources with a range of input magnitudes are added at empty sky positions.
We then run SExtractor with the same configuration file as for the catalogs and calculate the fraction of artificial sources recovered as a function of input magnitude.
This recovery fraction provides a sense as to the depth of each image.
Magnitudes corresponding to the 80\% recovery limit are listed in Table \ref{tab:photometry}.

We then use the 80\% completeness magnitude of the detection image for each LSS field to estimate stellar mass completeness limits.
In order to convert magnitude limits into stellar mass limits we make use of the FourStar Galaxy Evolution Survey \citep[ZFOURGE:][]{Straatman2016}.
For a given LSS field, we select all galaxies in ZFOURGE which have a magnitude that is within $\pm$0.1 mag of the 80\% limiting magnitude in the corresponding detection bandpass (for cases where the detection image is a stack of multiple images we use inverse-variance weighted mean magnitudes from ZFOURGE).
This is a representative sample of galaxies that are at the threshold of detection in our data, or to put it more precisely, these represent the galaxies that would be detected at a rate of 80\%.
Plotting stellar mass versus redshift for this sample we adopt the upper 95$^{\mathrm{th}}$ percentile as a function of redshift as the limiting stellar mass M$_{*, \mathrm{lim}}$.
Despite the fact that ZFOURGE is a K$_{\mathrm{s}}$-band selected survey it is complete to lower stellar masses.
We estimate the ZFOURGE survey to be between 0.2-1.5 dex deeper in terms of stellar mass completeness at the redshifts of each respective ORELSE large scale structure.

We also consider mass completeness limits more appropriate for galaxies dominated by old stellar populations which have elevated mass-to-light ratios.
These limits are derived from a single-burst BC03 stellar population with $z_{\mathrm{form}} = 5$ with no dust attenuation. 
At all $z < z_{\mathrm{form}}$ the model is scaled to the limiting magnitude of the detection bandpass and its stellar mass is recorded as M$_{\mathrm{*, ssp}}$.
Throughout this analysis we adopt the former (M$_{*, \mathrm{lim}}$) as the completeness limit for the total as well as star-forming galaxy populations and the latter (M$_{\mathrm{*, ssp}}$) as the completeness limit for the quiescent population.
All stellar mass completeness limits are listed in Table \ref{tab:masslim}.

\subsection{Classifying Star-Forming and Quiescent Galaxies}
\label{sec:uvj}

It has been shown that a simple combination of rest-frame broadband colours in the $U$, $V$, and $J$ filters can be an effective way of classifying galaxies as actively star-forming vs. quenched \citep[e.g.][]{Wuyts2007, Williams2009}.
The main advantage of this method is in its ability to decouple dust-reddened star-forming galaxies from quiescent galaxies in the $U - V$ versus $V - J$ colour space, thus allowing for a less biased selection of active and passive galaxies.

We make use of {\tt EAZY} in estimating rest-frame fluxes of galaxies.
For each galaxy the best-fit \zphot\ from the initial run of {\tt EAZY} (or \zspec\ when available) is used to identify which observed-frame photometry is nearby a given rest-frame filter.
{\tt EAZY} then fits a SED to these identified photometry and interpolates accordingly to extract a flux for the rest-frame filter.
This procedure is performed separately for each of the rest-frame $U$, $V$, and $J$ bands.

Figure \ref{fig:uvj} shows the distribution of ORELSE galaxies in the $UVJ$ diagram for our sample.
The boundaries delineating star-forming and quiescent are taken from \citet{Whitaker2011}; quiescent galaxies are selected with: $(U - V) > 1.3 \; \cap \; (V - J) < 1.6 \; \cap \; (U - V) > 0.88 (V - J) + 0.59$

% ~~~~~~~~~~~~~~~~~~~~~~~~~~~~~~~~~~~~~~~~~~~~~~~~~~~~~~~~~~~~~
% ~~~~~~~~~~~~~~~~~~~~~~~~~~~~~ FIGURE ~~~~~~~~~~~~~~~~~~~~~~~~~
% ~~~~~~~~~~~~~~~~~~~~~~~~~~~~~~~~~~~~~~~~~~~~~~~~~~~~~~~~~~~~~
\begin{figure*}
	\includegraphics[width=2\columnwidth]{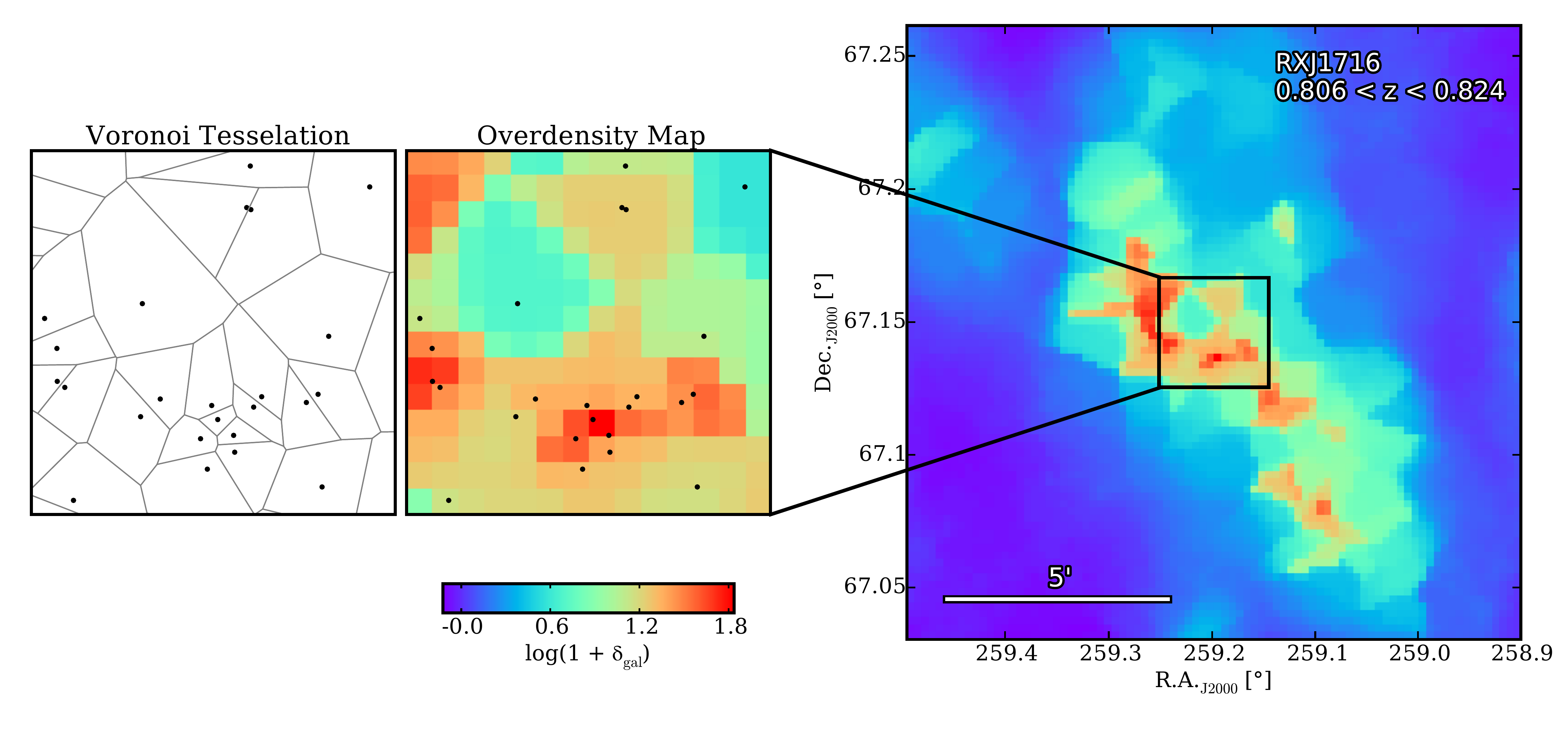}
	\caption{
	A brief illustration of how the Voronoi Monte-Carlo overdensity maps are constructed.
	For each LSS field, galaxies are separated into narrow redshift bins ($\Delta \mathrm{v} = \pm 1500$ km/s) across the broad redshift range of $0.55 < z < 1.3$.
	Photometric redshifts are used only when a spectroscopic redshift is not available.
	Here we show one such redshift slice bracketing the central redshift of the main overdensity in the RXJ 1716 field of view.
	A Voronoi tessellation is generated, separating galaxies into polygonal cells which represent the projected area that is nearest to the galaxy (left panel).
	Densities are calculated as the inverse of the cell area multiplied by the square of the angular diameter distance at the corresponding redshift.
	These densities are then projected onto a 2D pixel grid with a pixel scale of 75$\times$75 proper kpc.
	Finally, overdensities are calculated as $\mathrm{log}(1 + \delta_{\mathrm{gal}}) \equiv \mathrm{log}(1 + (\Sigma_{i, j} - \tilde{\Sigma}) / \tilde{\Sigma})$, where $\Sigma_{i,j}$ is the density of pixel ({\it i, j}) and $\tilde{\Sigma}$ is the median density (centre panel).
	This procedure is repeated 100 times where for each iteration photometric redshifts are randomly sampled based on the estimated \zphot\ uncertainties.
	The final overdensity map is computed by median-stacking the maps generated from all 100 iterations.
	%{\bf NOTE: I may make an updated version of this figure.}
	}
	\label{fig:voronoimap}
\end{figure*}

\subsection{Defining Local Environment}
\label{sec:voronoi_mc}

There exist a variety of quantitative metrics for defining the environment that a galaxy resides in.
In a qualitative sense these metrics can be viewed as falling into one of two broad categories: ``local environment'' which relates to small physical scales internal to the halo of a given galaxy and ``large-scale environment'' which relates to large physical scales external to the halo of a given galaxy \citep{Muldrew2012}.
For the purposes of this work we focus on a galaxy's local environment and its impact on the galaxy SMF.

We measure local environment for galaxies in our sample using a Voronoi Monte-Carlo algorithm which will be described in full detail in a future paper (Lemaux et al. in preparation).
An illustration of the approach described below is shown in Figure \ref{fig:voronoimap}.
The Voronoi tessellation has been shown to be one of the most accurate techniques for estimating the local density field in simulations \citep{Darvish2015}.
For each MC realization all objects that lack a high-quality \zspec\ are assigned a perturbed \zphot .
A perturbed \zphot\ for a given galaxy is calculated by randomly sampling a Gaussian with a mean and dispersion set to the original \zphot\ and 1$\sigma$  uncertainty, respectively.
All galaxies with a good use-flag are then divided into thin \z\ slices and a Voronoi tessellation is calculated on each slice.
These \z\ slices span $0.55 < z < 1.3$ and have widths set such that they encompass $\pm 1500$ km/s in velocity space from their central \z .
The areas of the Voronoi cells are then projected onto a two dimensional grid of $75 \times 75$ proper kpc pixels.
Local density is defined as the inverse of the cell area multiplied by the square of the angular diameter distance.
A final density map is computed by median combining the density maps from 100 Monte-Carlo realizations.
The local overdenisty at pixel $(i, j)$ is calculated as $\mathrm{log}(1 + \delta_{\mathrm{gal}}) \equiv \mathrm{log}(1 + (\Sigma_{i, j} - \tilde{\Sigma}) / \tilde{\Sigma})$, where $\tilde{\Sigma}$ is the median density of all pixels where the map is reliable (i.e., coverage in nearly all images and not near the edge of the detection image).
For reference, the largest (projected) environmental densities probed by our data reach as high as $\sim$100 galaxies/Mpc$^2$.
Note that this is based on the smallest Voronoi cells in the creation of the density maps and does not mean that we observe a single contiguous 1 Mpc$^2$ area containing 100 galaxies.

Note that true \zphot\ PDF of a galaxy is not always well-represented by a Gaussian distribution which could introduce a bias in the calculations described above.
To test this we have remeasured the density maps for one of our fields (RCS0224) using the \zphot\ PDFs output by {\tt EAZY} and directly compared the output overdensity values, pixel by pixel, to the fiducial case where we assume a Gaussian.
From this comparison we calculate that the median offset and NMAD scatter in $\mathrm{log}(1 + \delta_{\mathrm{gal}})$, between the Gaussian and formal PDF cases, is 0.005 dex and 0.18 dex respectively.
These values do not vary significantly with redshift.
Furthermore, these values are comparable to the level of variation seen when the density maps are recreated from a new 100-iteration MC calculation using the same (Gaussian) formalism.
Therefore, we conclude that (1) the assumption of Gaussianity of the \zphot\ PDF does not bias the construction of the Voronoi MC (over)density maps, and (2) the offset and scatter quoted above are representative of the natural level of deviation rooted in the Monte Carlo algorithm itself.

In order to test our ability to recover the true underlying density field, and thus the accuracy with which the galaxies are assigned densities from the Voronoi MC maps, we perform a series of tests on mock galaxy catalogs.
These mock catalogs were created to simulate three galaxy groups and one galaxy cluster at $z \sim 0.8$, similar to the SG0023 system studied in \citet{Lemaux2016arxiv} and ostensibly spanning the range of environments probed by ORELSE.
First, a sample of field galaxies were generated, randomly dispersed within a $\sim$15$\times$15\arcmin\ region between $z=0.7-0.9$.
Although this redshift range seems relatively narrow, it represents roughly $\pm 2 \sigma_{\Delta z / (1 + z_{\mathrm{spec}})} \times (1+z)$ from the central redshift and thus includes essentially all galaxies that would meaningfully contribute to the density map at $z \sim 0.8$.
Members of a galaxy group are inserted by randomly sampling from a 3D Gaussian of $\sigma = 0.33$ Mpc (proper) in all spatial dimensions with an additional scatter placed on the line-of-sight direction designed to mimic a velocity dispersion of 500 km/s.
The galaxy cluster is populated in the same way but instead sampled with $\sigma = 0.5$ Mpc and perturbed to mimic a velocity dispersion of 1000 km/s.
Luminosities are assigned to simulated galaxies by sampling from the rest-frame B-band luminosity function of \citet{Giallongo2005} at the appropriate redshifts.
For group and cluster galaxies the value of $M^*$ was set to be 0.25 and 0.5 mags brighter respectively in accordance with recent findings on the luminosity function in these types of environments \citep[e.g.][]{DePropris2013}.

% ~~~~~~~~~~~~~~~~~~~~~~~~~~~~~~~~~~~~~~~~~~~~~~~~~~~~~~~~~~~~~
% ~~~~~~~~~~~~~~~~~~~~~~~~~~~~~ FIGURE ~~~~~~~~~~~~~~~~~~~~~~~~~
% ~~~~~~~~~~~~~~~~~~~~~~~~~~~~~~~~~~~~~~~~~~~~~~~~~~~~~~~~~~~~~
\begin{figure*}
	\includegraphics[width=2\columnwidth]{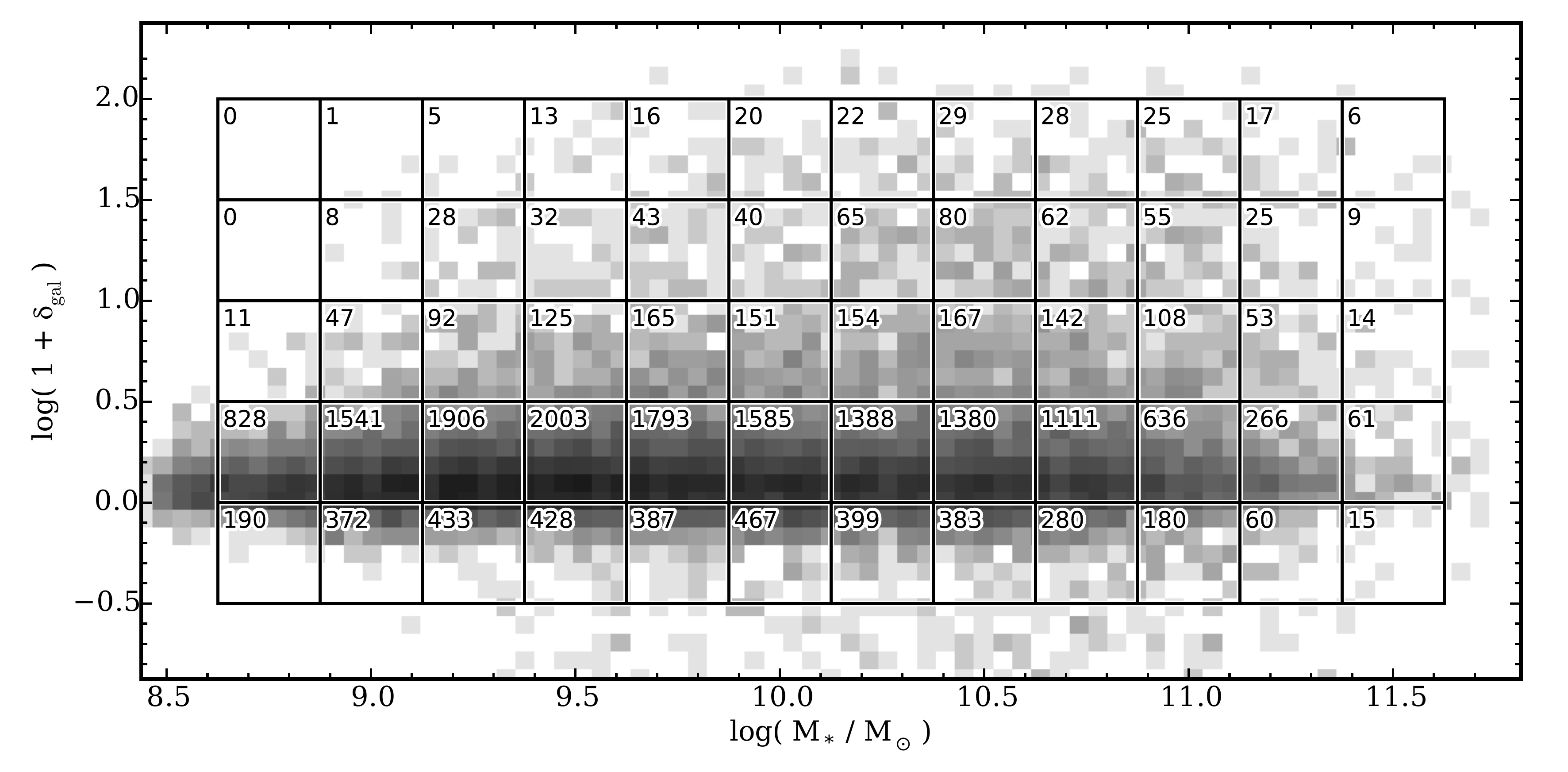}
	\caption{
	The distribution of galaxies in the overdensity versus stellar mass plane.
	The greyscale shows a logarithmic scaling of a 2D histogram of all galaxies at $0.55 < z < 1.3$ that are above our estimated stellar mass completeness limits (see Section \ref{sec:masslimits}).
	Vertical and horizontal lines trace out the binning scheme used to measure the galaxy SMFs in this analysis.
	The total number of galaxies in each bin is indicated in the top-left corner of each box.
	We refer the reader to Section \ref{sec:smf_construction} for a full description of how the SMF is constructed from this distribution.
	}
	\label{fig:overdensity_mass}
\end{figure*}

After creating mock catalogs from this process, we generate density and overdensity maps following the procedures described earlier, applying identical magnitude cuts.
No $k$-correction was applied to the rest-frame B-band magnitudes when translating them to observed-frame $i^{\prime}$ magnitudes as this correction is only $\sim$0.1 mags at these redshifts and varies weakly with galaxy type.
We test a set of scenarios with differing rates of spectroscopic coverage (fraction of all galaxies assumed to have spectroscopic redshifts) ranging from $3-78$\%.
Galaxies that do not have a spectroscopic redshift are assumed to have a photometric redshift with a precision of $0.03 \times (1+z)$ and a catastrophic outlier rate of 6\%; this is accomplished by perturbing their initial, true (redshift space) redshifts by a values consistent with these statistics.
We find that in a case with $\approx$20\%\ spectroscopic coverage, roughly 60\%\ and 80\%\ of galaxies lie within $\pm$0.25 dex  and $\pm$0.5 dex of their true overdensity respectively.
This level of spectroscopic completeness is roughly equal to that of all ORELSE fields considered in this study over the spectroscopic footprint and subject to the magnitude ranges used to generate density and overdensity maps.
Further, these recovery rates increase rapidly as the spectroscopic coverage increases, an increase which is observed in ORELSE with increasing $\log(1+\delta_{\mathrm{gal}})$ \citep[see][]{Lemaux2016arxiv, Shen2017arxiv}.
Therefore, we conclude that over the broad range of overdensities examined in this study (2.5 orders of magnitude, as discussed in the following section) we are able to quantify the environments of galaxies with sufficient accuracy.
A more detailed description and analysis of these tests will be presented in a future publication (Lemaux et al. in preparation).

% ~~~~~~~~~~~~~~~~~~~~~~~~~~~~~~~~~~~~~~~~~~~~~~~~~~~~~~~~~~~~~
% ~~~~~~~~~~~~~~~~~~~~~~~~~~~~~ FIGURE ~~~~~~~~~~~~~~~~~~~~~~~~~
% ~~~~~~~~~~~~~~~~~~~~~~~~~~~~~~~~~~~~~~~~~~~~~~~~~~~~~~~~~~~~~
\begin{figure*}
	\includegraphics[width=2\columnwidth]{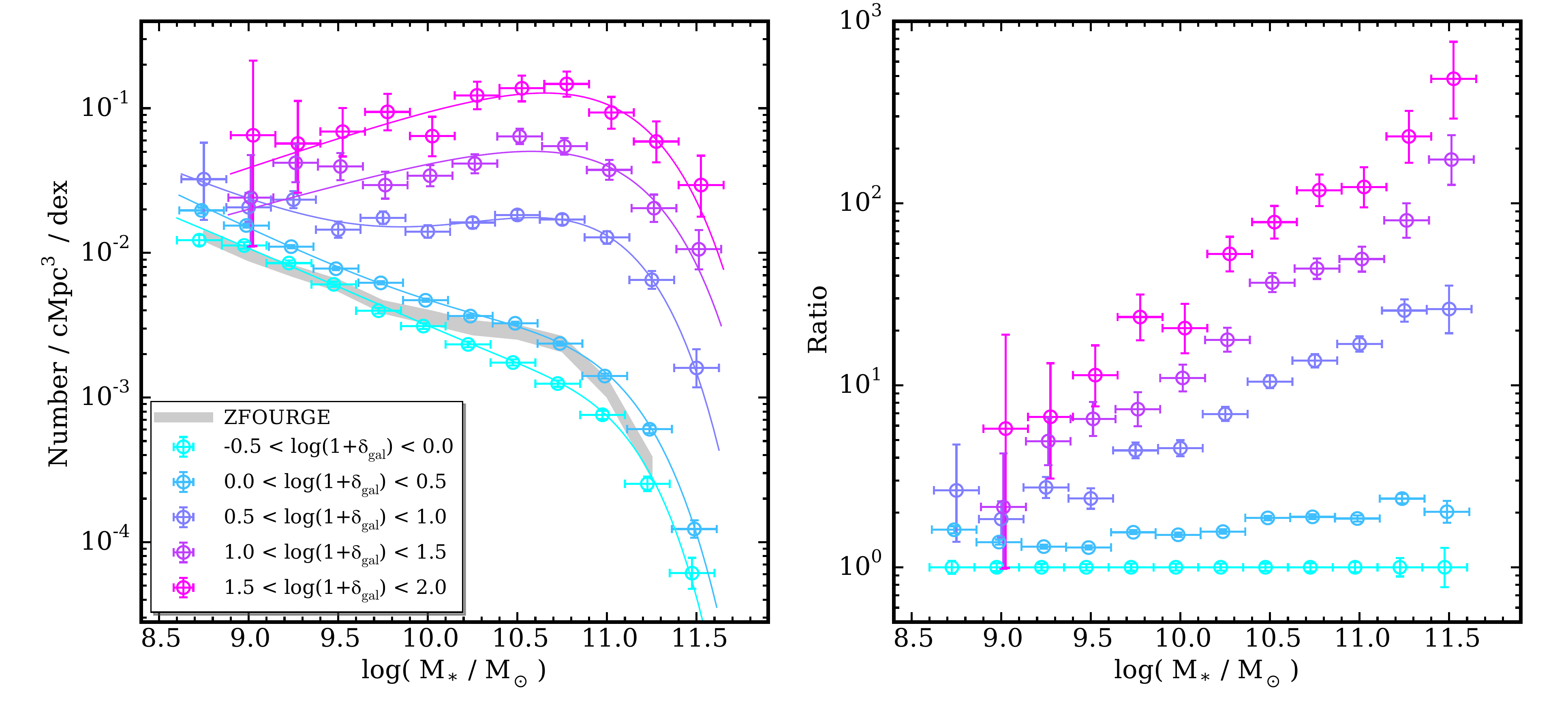}
	\caption{
	In the left panel are shown the galaxy stellar mass functions (normalized by comoving volume) for all galaxies at $0.55 < z < 1.3$ in five bins of local environmental overdensity.
    Curves showing the best-fitting Schechter function (single or double) which yields the lower BIC as shown in Table \ref{tab:params}.
	These measurements span a broad dynamic range of environments from the average densities hosting field galaxies to the central cores of massive galaxy clusters.
	For reference we plot the field galaxy stellar mass function at $0.5 < z < 1.25$ as measured by ZFOURGE \citep{Tomczak2014} which shows remarkable consistency with a combination of our two lowest-density SMFs as expected.
	It can be clearly seen that as environmental density increases there is a smooth, continuous increase in the relative numbers of higher- to lower-mass galaxies.
	This behavior is elucidated in the right-hand panel which plots the ratio of each SMF relative to that of the lowest density bin.
	This type of trend is consistent with the picture that dense environments preferentially destroy lower-mass galaxies and/or promote the growth of higher-mass galaxies, though it is likely that these mechanisms are interdependent.
%	These results may also shed some light as to the origin of the double-Schechter behavior of the galaxy SMF as measured from prominent galaxy legacy surveys \citep[e.g.][]{Ilbert2013, Tomczak2014}.
%	could be the result of averaging over a range of environments, the denser of which provide a mild excess at the high-mass end producing the distinctive double-Schechter behavior.
	}
	\label{fig:mfs_vor}
\end{figure*}

\subsection{Measuring the SMF in Different Environments}
\label{sec:smf_construction}

To create samples of galaxies occupying different local environments we define five bins of local overdensity between -0.5 $\leq$ log(1+$\delta_{\mathrm{gal}}$) $\leq$ 2 each of width 0.5 dex.
For each of these overdensity bins we construct the galaxy SMF by counting numbers of galaxies in stellar mass bins of width 0.25 dex.
Figure \ref{fig:overdensity_mass} shows an illustration of this binning scheme.
Because of the heterogeneity of our catalogs and the subsequent variety of mass completeness limits we employ a methodology that effectively mimics the $1 / V_{\mathrm{max}}$ technique \citep{Avni1980}.
In brief the concept of this approach is to evaluate the galaxy number density separately for each stellar mass bin, only considering the volume for which the catalogs are mass-complete.

\begin{figure*}
	\includegraphics[width=2\columnwidth]{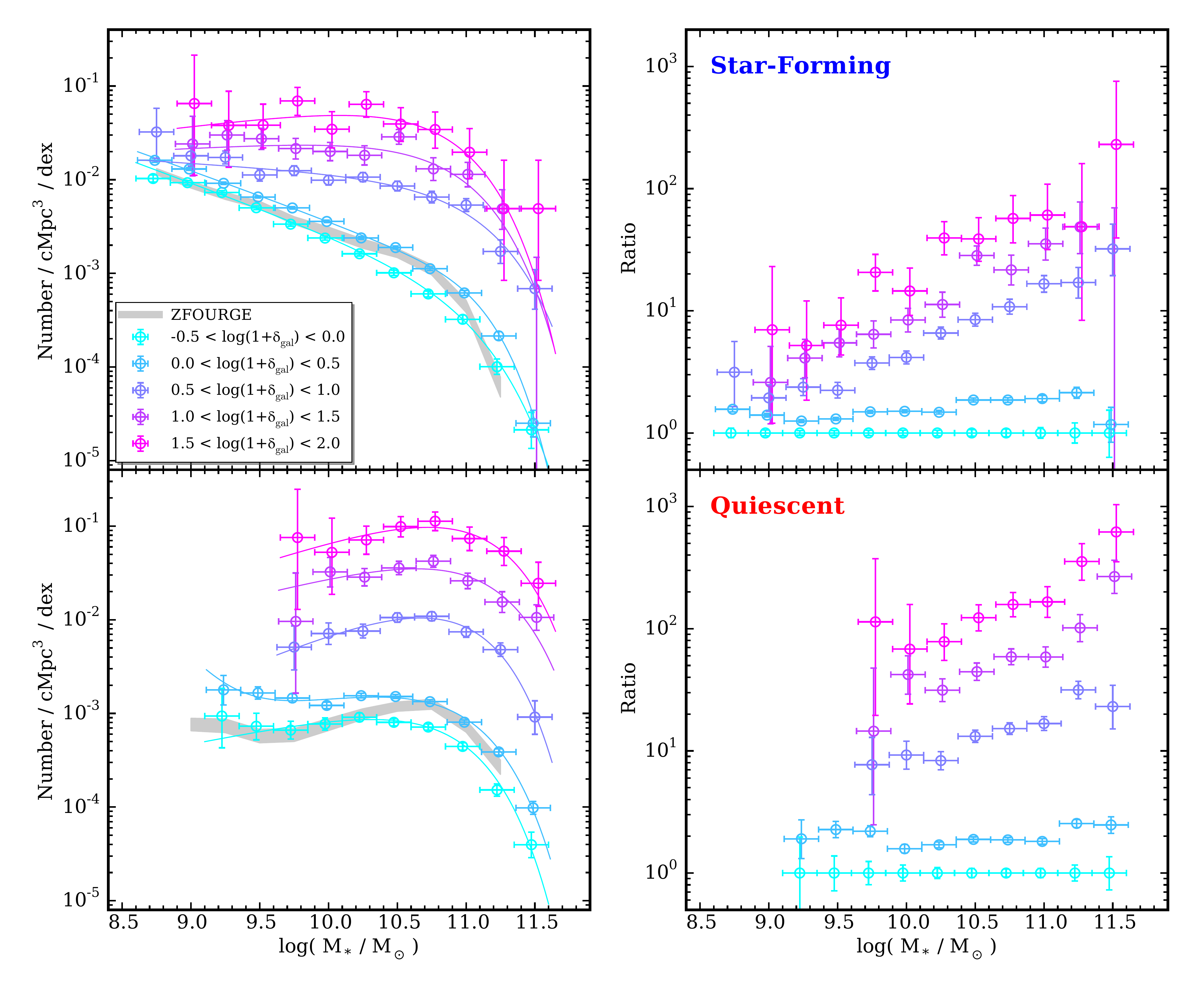}
	\caption{
	Galaxy stellar mass functions split into star-forming (top panels) and quiescent (bottom panels) subpopulations based on rest-frame broadband colours.
	Symbols and curves are the same as described in Figure \ref{fig:mfs_vor}.
	Similar to the case for all galaxies, both the star-forming and quiescent subsamples exhibit an environmental dependence on the shape of their respective SMFs.
	Although there is still general consistency between ZFOURGE and our measurements in field environments, there are some differences that jump out.
	First, for quiescent galaxies the linear combination of the two lowest density SMFs overestimates the number densities of $\lesssim 10^{10}$ \msol\ galaxies relative to ZFOURGE by about a factor of 2$\times$.
	Second, for star-forming galaxies there is similarly a factor of $\sim$2$\times$ excess in our measured number densities of $\gtrsim 10^{11.2}$ \msol\ which may be the result of Eddington bias \citep{Eddington1913}.
    }
	\label{fig:mfs_vor_sfqu}
\end{figure*}

We perform this by stepping through each of the \z\ slices used to construct the overdensity maps discussed in Section \ref{sec:voronoi_mc} and identify the spatial area and corresponding volume associated with each overdensity bin.
If a given stellar mass bin is partially below the limiting stellar mass at the central redshift of the slice then the volume is ignored and galaxies are not counted.
Otherwise all galaxies that fall within the \z\ bounds of the slice (\zspec\ when available) are counted towards the stellar mass function of the corresponding mass-overdenisty bin.
As a result of this approach, some stellar mass bins will count galaxies from smaller volumes than others based on whether they lie above/below the mass-completeness limit in a given \z\ slice for different numbers of fields across the full \z\ range $0.55 < z < 1.3$.
Each stellar mass bin is normalized by the total comoving volume probed for that bin.
In Figures \ref{fig:mfs_vor} and \ref{fig:mfs_vor_sfqu} we show our measurements of the SMF of all galaxies as well as the star-forming \& quiescent components respectively.

To test the stability of the SMFs measured in this way we perform a jackknife analysis on the eight ORELSE fields used in this study.
We do this by remeasuring all SMFs using every permutation of eight, seven, six ... one field(s), amounting to $2^8 - 1 = 255$ realizations in total.
With these measurements we calculate the 1$\sigma$ scatter for each data point of each SMF in Figures \ref{fig:mfs_vor} and \ref{fig:mfs_vor_sfqu} respectively.
We find that the median 1$\sigma$ scatter per data point for the Total mass functions from the lowest to the highest overdensity bins is 0.05, 0.04, 0.07, 0.12, and 0.12 dex respectively.
For the Star-Forming SMFs this scatter is 0.06, 0.04, 0.08, 0.14, and 0.18 dex respectively and for the Quiescent SMFs it is 0.09, 0.07, 0.09, 0.12, and 0.13 dex respectively.
Next, we refit Schechter functions to each jackknife realization of each SMF.
In general, the 1$\sigma$ scatter for each best-fit Schechter parameter across all jackknife realizations is on average only 1.37$\times$ larger than the formal uncertainties quoted in Table \ref{tab:params}.
Therefore, we conclude that the overall measurements of the SMFs are largely stable, despite some minor variation among the eight ORELSE fields used in this study.

% ~~~~~~~~~~~~~~~~~~~~~~~~~~~~~~~~~~~~~~~~~~~~~~~~~~~~~~~~~~~~~
% ~~~~~~~~~~~~~~~~~~~~~~~~~~~~~ TABLE ~~~~~~~~~~~~~~~~~~~~~~~~~~
% ~~~~~~~~~~~~~~~~~~~~~~~~~~~~~~~~~~~~~~~~~~~~~~~~~~~~~~~~~~~~~

\begin{table*}
	\begin{center}
	\caption{Schechter Parameters}
	\label{tab:params}

	Total \\[0.5mm]
	\begin{tabular}{ccccccc}

	\hline \\[-5mm]
	\hline \\[-3mm]

	Overdensity bin & log($M^*$) & $\alpha_1$ & $\Phi^*_1$ & $\alpha_2$ & $\Phi^*_2$ & BIC \\
	 & [$M_{\odot}$] & & [$10^{-3}$ Mpc$^{-3}$] & & [$10^{-3}$ Mpc$^{-3}$] \\
	\hline \\[-2mm]

    -0.5 $<$ log(1+$\delta_{\mathrm{gal}}$) $<$ 0.0 & $11.17\pm0.07$ & $-1.46\pm0.03$ & $ 0.11\pm0.02$ & ... & ... & 44.3  \\
    -0.5 $<$ log(1+$\delta_{\mathrm{gal}}$) $<$ 0.0 & $10.77\pm0.11$ & $-0.14\pm0.47$ & $ 0.21\pm0.04$ & $-1.52\pm0.06$ & $ 0.14\pm0.05$ & 27.4  \\[-0.5mm]
    \hline
    0.0 $<$ log(1+$\delta_{\mathrm{gal}}$) $<$ 0.5 & $11.20\pm0.08$ & $-1.38\pm0.04$ & $ 0.21\pm0.04$ & ... & ...  & 161.8  \\
    0.0 $<$ log(1+$\delta_{\mathrm{gal}}$) $<$ 0.5 & $10.87\pm0.05$ & $-0.59\pm0.20$ & $ 0.45\pm0.04$ & $-1.60\pm0.07$ & $ 0.12\pm0.04$ & 25.6  \\[-0.5mm]
    \hline
    0.5 $<$ log(1+$\delta_{\mathrm{gal}}$) $<$ 1.0 & $11.07\pm0.08$ & $-0.85\pm0.07$ & $ 2.99\pm0.51$ & ... & ... & 36.6  \\
    0.5 $<$ log(1+$\delta_{\mathrm{gal}}$) $<$ 1.0 & $10.87\pm0.08$ & $-0.36\pm0.27$ & $ 4.34\pm0.40$ & $-1.52\pm0.32$ & $ 0.25\pm0.37$ & 19.0  \\[-0.5mm]
    \hline
    1.0 $<$ log(1+$\delta_{\mathrm{gal}}$) $<$ 1.5 & $11.04\pm0.10$ & $-0.65\pm0.10$ & $11.16\pm2.03$ & ... & ... & 21.1  \\
    1.0 $<$ log(1+$\delta_{\mathrm{gal}}$) $<$ 1.5 & $10.93\pm0.16$ & $-0.43\pm0.45$ & $13.47\pm2.70$ & $-1.66\pm1.58$ & $ 0.17\pm1.23$ & 22.8  \\[-0.5mm]
    \hline
    1.5 $<$ log(1+$\delta_{\mathrm{gal}}$) $<$ 2.0 & $11.03\pm0.09$ & $-0.58\pm0.10$ & $30.35\pm4.98$ & ... & ... & 11.0  \\
    1.5 $<$ log(1+$\delta_{\mathrm{gal}}$) $<$ 2.0 & $10.98\pm0.15$ & $-0.49\pm0.31$ & $33.01\pm7.78$ & $-2.00\pm3.54$ & $ 0.05\pm0.76$ & 15.8  \\[-0.5mm]
	\hline \\[-2.5mm]

	\multicolumn{7}{c}{Star-Forming} \\[-0.5mm]
	\hline \\[-5mm]
	\hline \\[-3mm]

    -0.5 $<$ log(1+$\delta_{\mathrm{gal}}$) $<$ 0.0 & $11.03\pm0.05$ & $-1.54\pm0.02$ & $ 0.08\pm0.01$ & ... & ... & 19.0  \\
    -0.5 $<$ log(1+$\delta_{\mathrm{gal}}$) $<$ 0.0 & $10.68\pm0.12$ & $ 0.49\pm0.52$ & $ 0.06\pm0.02$ & $-1.50\pm0.04$ & $ 0.15\pm0.04$ & 22.5  \\[-0.5mm]
    \hline
    0.0 $<$ log(1+$\delta_{\mathrm{gal}}$) $<$ 0.5 & $11.06\pm0.04$ & $-1.48\pm0.02$ & $ 0.13\pm0.02$ & ... & ... & 34.7  \\
    0.0 $<$ log(1+$\delta_{\mathrm{gal}}$) $<$ 0.5 & $10.76\pm0.08$ & $-0.21\pm0.37$ & $ 0.17\pm0.03$ & $-1.52\pm0.04$ & $ 0.17\pm0.04$ & 22.9  \\[-0.5mm]
    \hline
    0.5 $<$ log(1+$\delta_{\mathrm{gal}}$) $<$ 1.0 & $11.09\pm0.08$ & $-1.09\pm0.05$ & $ 1.04\pm0.18$ & ... & ... & 16.2  \\
    0.5 $<$ log(1+$\delta_{\mathrm{gal}}$) $<$ 1.0 & $10.98\pm0.11$ & $-0.93\pm0.21$ & $ 1.40\pm0.35$ & $-2.13\pm1.11$ & $ 0.01\pm0.04$ & 17.6  \\[-0.5mm]
    \hline
    1.0 $<$ log(1+$\delta_{\mathrm{gal}}$) $<$ 1.5 & $10.91\pm0.12$ & $-0.92\pm0.12$ & $ 3.36\pm0.94$ & ... & ... & 14.8  \\
    1.0 $<$ log(1+$\delta_{\mathrm{gal}}$) $<$ 1.5 & $10.70\pm0.27$ & $-0.09\pm1.44$ & $ 3.98\pm2.48$ & $-1.26\pm0.63$ & $ 1.23\pm3.01$ & 17.7  \\[-0.5mm]
    \hline
    1.5 $<$ log(1+$\delta_{\mathrm{gal}}$) $<$ 2.0 & $10.83\pm0.15$ & $-0.82\pm0.16$ & $ 8.57\pm2.70$ & ... & ... & 11.3  \\
    %%%  from p0=[10.9, 0.0, 0.0, 0.01, 0.1]
    1.5 $<$ log(1+$\delta_{\mathrm{gal}}$) $<$ 2.0 & $10.81\pm0.18$ & $-0.81\pm0.20$ & $ 8.80\pm3.36$ & $-15.45\pm11.64$ & 2e-26 $\pm$ 6e-22 & 16.1  \\[-0.5mm]
	\hline \\[-2.5mm]

	\multicolumn{7}{c}{Quiescent} \\[-0.5mm]
	\hline \\[-5mm]
	\hline \\[-3mm]

    -0.5 $<$ log(1+$\delta_{\mathrm{gal}}$) $<$ 0.0 & $10.84\pm0.03$ & $-0.71\pm0.05$ & $ 0.18\pm0.01$ & ... & ... & 10.3  \\
    -0.5 $<$ log(1+$\delta_{\mathrm{gal}}$) $<$ 0.0 & $10.84\pm0.04$ & $-0.68\pm0.08$ & $ 0.18\pm0.02$ & $-4.20\pm5.22$ & 4e-7 $\pm$ 8e-6 & 14.4  \\[-0.5mm]
    \hline
    0.0 $<$ log(1+$\delta_{\mathrm{gal}}$) $<$ 0.5 & $10.93\pm0.04$ & $-0.78\pm0.06$ & $ 0.29\pm0.03$ & ... & ... & 22.2  \\
    0.0 $<$ log(1+$\delta_{\mathrm{gal}}$) $<$ 0.5 & $10.91\pm0.05$ & $-0.71\pm0.13$ & $ 0.31\pm0.04$ & $-2.78\pm2.14$ & $ 0.0001\pm0.001$ & 21.7  \\[-0.5mm]
    \hline
    0.5 $<$ log(1+$\delta_{\mathrm{gal}}$) $<$ 1.0 & $10.89\pm0.05$ & $-0.39\pm0.10$ & $ 2.82\pm0.23$ & ... & ... &  8.4  \\
    %%%  from p0=[10.83, -1.0, -1.0, 0.01, 1e-05]
    0.5 $<$ log(1+$\delta_{\mathrm{gal}}$) $<$ 1.0 & $10.88\pm0.08$ & $-0.38\pm0.23$ & $ 2.83\pm0.35$ & $-3.61\pm54.15$ & $ 0.0006\pm0.01$ & 12.6  \\[-0.5mm]
    \hline
    1.0 $<$ log(1+$\delta_{\mathrm{gal}}$) $<$ 1.5 & $11.07\pm0.13$ & $-0.63\pm0.20$ & $ 7.97\pm2.01$ & ... & ... & 11.3  \\
    %%%  from p0=[10.9, -1, 10, 0.01, 1e-08]
    1.0 $<$ log(1+$\delta_{\mathrm{gal}}$) $<$ 1.5 & $10.69\pm0.18$ & $-0.11\pm0.35$ & $11.59\pm1.38$ & $ 4.65\pm1.23$ & $ 0.02\pm0.04$ & 12.4  \\[-0.5mm]
    \hline
    1.5 $<$ log(1+$\delta_{\mathrm{gal}}$) $<$ 2.0 & $11.04\pm0.09$ & $-0.52\pm0.15$ & $24.18\pm3.89$ & ... & ... &  7.3  \\
    %%%  from p0=[10.83, -1.0, -1.0, 0.01, 1e-05]
    1.5 $<$ log(1+$\delta_{\mathrm{gal}}$) $<$ 2.0 & $10.56\pm0.09$ & $ 0.25\pm0.26$ & $31.44\pm1.53$ & $ 4.87\pm1.09$ & $ 0.05\pm0.09$ & 11.0  \\
	\hline \\[-2.5mm]

	\end{tabular}
	\end{center}
\end{table*}

% ~~~~~~~~~~~~~~~~~~~~~~~~~~~~~~~~~~~~~~~~~~~~~~~~~~~~~~~~~~~~~
% ~~~~~~~~~~~~~~~~~~~~~~~~~~~~~ FIGURE ~~~~~~~~~~~~~~~~~~~~~~~~~
% ~~~~~~~~~~~~~~~~~~~~~~~~~~~~~~~~~~~~~~~~~~~~~~~~~~~~~~~~~~~~~

% ~~~~~~~~~~~~~~~~~~~~~~~~~~~~~~~~~~~~~~~~~~~~~~~~~~~~~~~~~~~~~
% ~~~~~~~~~~~~~~~~~~~~~~~~~~~~~ FIGURE ~~~~~~~~~~~~~~~~~~~~~~~~~
% ~~~~~~~~~~~~~~~~~~~~~~~~~~~~~~~~~~~~~~~~~~~~~~~~~~~~~~~~~~~~~
\begin{figure*}
	\includegraphics[width=2\columnwidth]{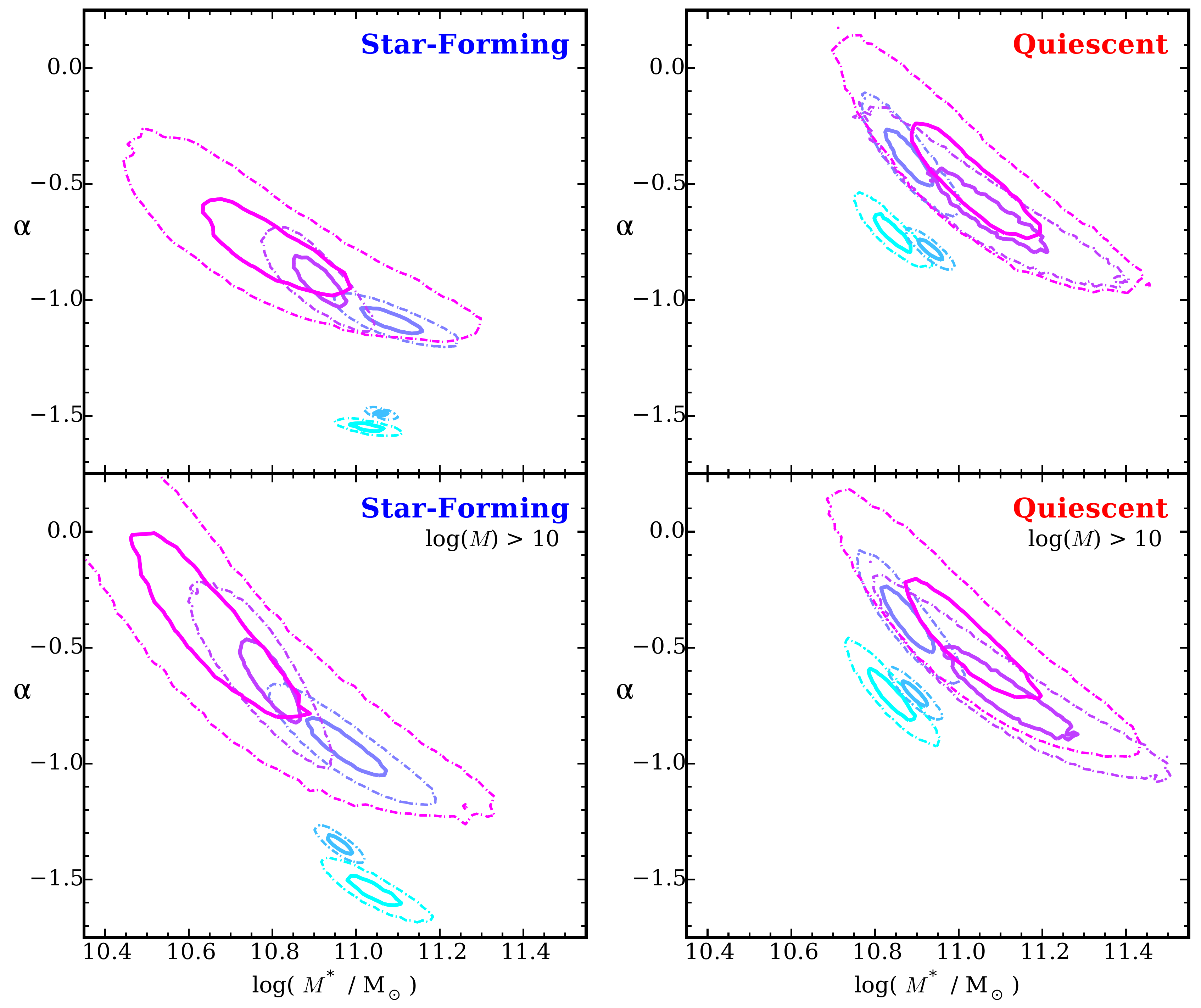}
	\caption{
	Likelihood contours for the single-Schechter parameters of the star-forming ({\it left}) and quiescent ({\it right}) SMFs.
	Despite the fact that some SMFs favor the double-Schechter function \citep[e.g.][]{Drory2009, Muzzin2013a, Tomczak2014, Mortlock2015}, we exclusively use the single-Schechter function here in order to perform a self-consistent comparison.
	Colours correspond to the same overdensity bins as in previous figures.
	Solid and dot-dashed lines indicate the 1$\sigma$ and 2$\sigma$ confidence levels respectively.
	For both the star-forming and quiescent subsamples we see a clear environmental dependence in the Schechter parameters.
	This feature persists even if the fitting is restricted to stellar masses $>$10$^{10}$ \msol\ as shown in the bottom panels.
	In particular the difference is roughly bimodal separating into loci for low- and high-density SMFs.
	This suggests that the transformation in the SMF begins to occur in group-scale environments where the local galaxy density is $\gtrsim3\times$ that of the field.
	}
	\label{fig:schechter_params}
\end{figure*}

\section{Results}
\label{sec:results}

\subsection{Dependence of the SMF on Local Environment}
\label{sec:smfs_tot}

We find that the shape of the galaxy stellar mass function at $0.55 < z < 1.3$ is a strong function of local environment, echoing similar works at these \z s \citep{Bolzonella2010, Vulcani2012, vanderBurg2013, Mortlock2015, Davidzon2016}.
In particular there is a continual and gradual increase in the relative number of high-mass to low-mass galaxies as environmental density increases.
This is most clearly illustrated in the right-hand panel of Figure \ref{fig:mfs_vor} which plots the ratio of each SMF relative that of the lowest density bin.
This ratio tends to follow an uninterrupted log-linear relationship with stellar mass becoming gradually steeper towards higher overdensities.
For every order of magnitude increase in stellar mass there is roughly a factor of 1.2, 3.0, 5.3, and 6.2 increase in the number of galaxies relative to the SMF of the lowest density bin in each of the higher overdensity bins respectively.

\renewcommand{\thefootnote}{\fnsymbol{footnote}}
Two possible explanations of this behavior are that regions of higher local overdensity ($a$) destroy lower mass galaxies at a quicker rate than they are added and/or ($b$) promote the growth of higher mass galaxies.
The former point is consistent with past studies that have indeed found that lower-mass galaxies are more likely to be ``destroyed'' via merging.
\citet{Leja2015} show that merging leads to a destruction rate of galaxies that is inversely proportional to stellar mass and a growth rate that is proportional with stellar mass within the semi-analytic model (SAM) of \citet{Guo2013b}.
Using a combined morphological and galaxy-pair counting analysis of the COSMOS legacy survey, \citet{Lotz2011} infer that the volume-averaged minor merger rate is 3x that of the major merger rate.
Furthermore, merger rates in galaxy group/cluster environments have been estimated to be 3-4$\times$ greater relative to lower-density environments \citep{Lin2010, Kampczyk2013}\footnote[3]{This result is likely driven by group-like environments which dominate the high-density sample investigated in these studies} which bolsters this picture.
\renewcommand{\thefootnote}{\arabic{footnote}}

Lastly, our measurements probe to low enough stellar masses to reveal, in some cases, a clear departure from standard single-Schechter behavior \citep{Schechter1976}.
However, it is not immediately clear that a more complex model is necessary in all cases, particularly for the higher density SMFs where uncertainties can be relatively large.
To test this we fit all SMFs with both single- and double-Schechter functions, respectively defined as:

\begin{equation}
\Phi_{\rm{single}} (M) = \;\; \mathrm{ln} (10) \; \Phi^* \left( \frac{M}{M^*} \right)^{\alpha + 1} \mathrm{exp} \left( -\frac{M}{M^*} \right)
\end{equation}

\begin{equation}
\begin{split}
\Phi_{\rm{double}} (M) =& \;\; \mathrm{ln} (10) \; \mathrm{exp} \left( -\frac{M}{M^*} \right) \times \\
&\left[ \Phi^*_1 \left( \frac{M}{M^*} \right)^{\alpha_1 + 1}   +   \Phi^*_2 \left( \frac{M}{M^*} \right)^{\alpha_2 + 1}   \right]
\end{split}
\end{equation}

\noindent
where \mstar\ is the characteristic turnover mass, ($\alpha$, $\alpha_1$, $\alpha_2$) are the slopes and ($\Phi^*$, $\Phi^*_1$, $\Phi^*_2$) are the normalizations.  
Next, for each fit we calculate the Bayesian Information Criterion \citep[BIC:][]{Schwarz1978} defined as:

\begin{equation}
{\rm BIC} = -2 \; {\rm ln}(L) + k \; {\rm ln}(n)
\end{equation}

\noindent
where $L$ is the maximum likelihood of the best-fit model, $k$ is the number of free parameters in the model, and $n$ is the number of measurements.
This statistic is designed to compare models with different numbers of free parameters and appropriately assess whether the improvement to the goodness of fit justifies an increase in the dimensionality of the model.
The model yielding the lower BIC is generally the favorable choice.
Best-fit Schechter parameterizations and their corresponding BIC values are shown in Table \ref{tab:params}.

% ~~~~~~~~~~~~~~~~~~~~~~~~~~~~~~~~~~~~~~~~~~~~~~~~~~~~~~~~~~~~~
% ~~~~~~~~~~~~~~~~~~~~~~~~~~~~~ FIGURE ~~~~~~~~~~~~~~~~~~~~~~~~~
% ~~~~~~~~~~~~~~~~~~~~~~~~~~~~~~~~~~~~~~~~~~~~~~~~~~~~~~~~~~~~~
\begin{figure*}
	\includegraphics[width=2\columnwidth]{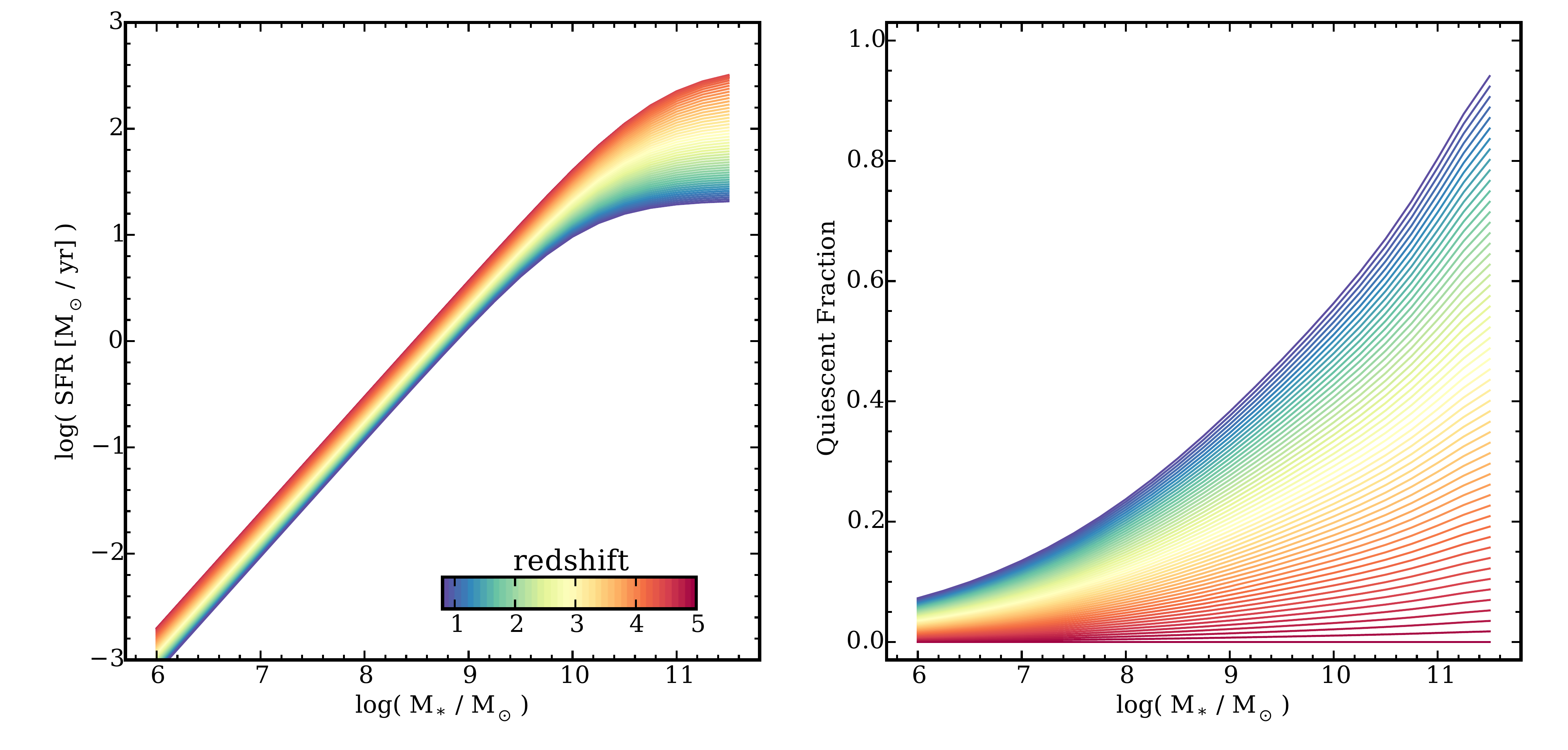}
	\caption{
	Left: Star-formation vs. stellar mass relations as a function of redshift used to construct star-formation histories for galaxies in the model described in Section \ref{sec:model}.
	These relations are taken from the analytical parameterizations presented in Equations 2 and 4 of \citet{Tomczak2016}.
	At each 100 Myr time-step between $z_{\rm{start}}$=5 and $z_{\rm{final}}$=0.8 non-quenched galaxies grow based on their instantaneous SFR and are reassigned a new SFR based on where they fall on the SFR-M$_*$ relation of the subsequent time-step.
	Right: An illustration of the prescribed evolution for the quiescent fraction vs. stellar mass of simulated galaxies as a function of redshift.
	For this prescription we adopt two boundary conditions: (1) all galaxies begin as star-forming at the onset of the simulation $z_{\rm{start}}$=5, and (2) the quiescent fraction at fixed stellar mass at the end of the simulation ($z_{\rm{final}}$=0.8) must match the observed quiescent fraction of the densest environment in ORELSE.
	At each time-step of the simulation, a certain fraction of galaxies at every stellar mass are quenched at random in order to gradually transition the quiescent fraction between these boundary conditions, as shown.
	}
	\label{fig:sfrmass_fquiescent_simulation}
\end{figure*}

% ~~~~~~~~~~~~~~~~~~~~~~~~~~~~~~~~~~~~~~~~~~~~~~~~~~~~~~~~~~~~~
% ~~~~~~~~~~~~~~~~~~~~~~~~~~~~~ FIGURE ~~~~~~~~~~~~~~~~~~~~~~~~~
% ~~~~~~~~~~~~~~~~~~~~~~~~~~~~~~~~~~~~~~~~~~~~~~~~~~~~~~~~~~~~~
\begin{figure}
	\includegraphics[width=\columnwidth]{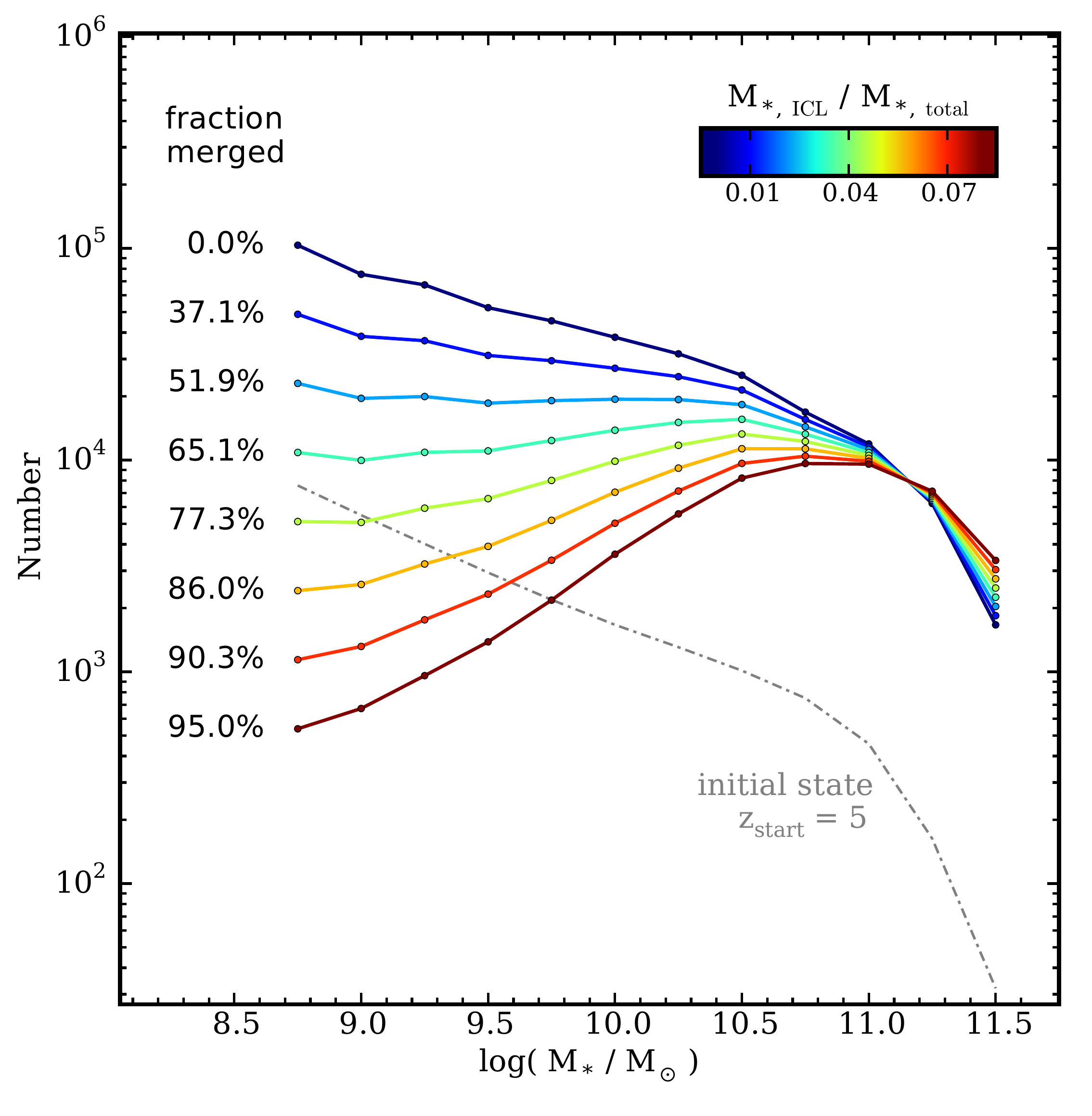}
	\caption{
	Simulated stellar mass functions from the semi-empirical model described in Section \ref{sec:model}.
	Briefly, the simulation begins at $z_{\rm{start}}$=5 with $\approx$10$^6$ galaxies having stellar masses down to $10^6$ \msol\ distributed in accordance with the measured SMF of the field.
	The simulation then proceeds in 100 Myr time increments down to $z_{\rm{final}}$=0.8 allowing galaxies to grow via star-formation, quench, and merge.
	The grey dot-dash line shows the assumed initial state at $z_{\rm{start}}$ and coloured solid lines show the resultant SMF at $z_{\rm{final}}$ for different values of $f_{\rm{merged}}$ (fraction of galaxies merged between $z_{\rm{start}}$ and $z_{\rm{final}}$).
	The colour-scale corresponds to the fraction of stellar mass in the ICL at $z_{\rm{final}}$ relative to total ($M_{*,\,\mathrm{ICL}} / M_{*,\,\mathrm{total}}$) estimated from the model based on assuming 30\%\ mass loss per merger.
	}
	\label{fig:merger_simulation}
\end{figure}

% ~~~~~~~~~~~~~~~~~~~~~~~~~~~~~~~~~~~~~~~~~~~~~~~~~~~~~~~~~~~~~
% ~~~~~~~~~~~~~~~~~~~~~~~~~~~~~ FIGURE ~~~~~~~~~~~~~~~~~~~~~~~~~
% ~~~~~~~~~~~~~~~~~~~~~~~~~~~~~~~~~~~~~~~~~~~~~~~~~~~~~~~~~~~~~
\begin{figure*}
	\includegraphics[width=2\columnwidth]{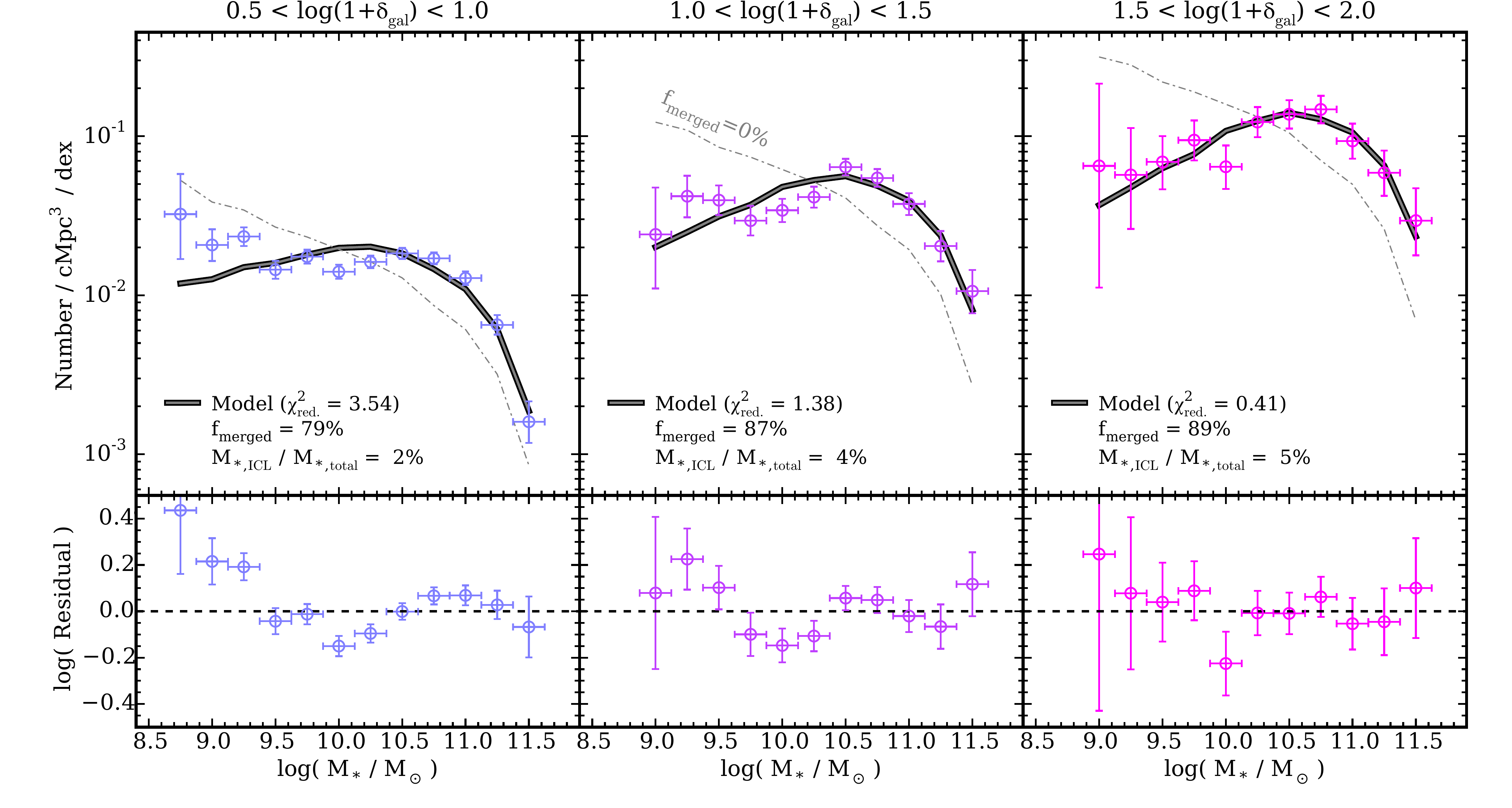}
	\caption{
	Thick solid lines show best-fits of our semi-empirical model to each of the measured SMFs in our three highest overdensity bins, with residuals plotted in the bottom panels.
	The grey dot-dashed curves show the version of the model in which no galaxy-galaxy mergers are included ($f_{\rm{merged}} = 0$\%).
	Despite the simplicity of the model it is able to reproduce the shape of each measured SMF reasonably well.
	In the bottom-left corners of the top panels we indicate the reduced $\chi^2$, the value of $f_{\rm{merged}}$, and the inferred stellar mass fractions of the ICL from the best-fit model.
	In general, our model shows that (1) galaxy mergers are necessarily important in shaping the SMF and (2) most of these mergers are likely to have occurred in environments of intermediate density such as galaxy groups.
	}
	\label{fig:merger_simulation_results}
\end{figure*}

\subsection{Star-Forming and Quiescent SMFs}

We further split the total galaxy SMF of each overdensity bin into its star-forming and quiescent components in Figure \ref{fig:mfs_vor_sfqu}.
Several recent studies have found that, similar to the total galaxy population, the SMFs of star-forming and quiescent galaxies also exhibit an upturn at $\lesssim 10^{9.5}$ \msol\ at the redshifts examined here \citep[e.g.][]{Drory2009, Muzzin2013a, Tomczak2014, Mortlock2015}.
As described in the previous section, we test for the presence of this feature in our measurements by comparing the Bayesian Information Criteria between the best-fit single- versus double-Schechter function.
From these comparisons we find that only the low-density 0$<$log(1+$\delta_{\mathrm{gal}}$)$<$0.5 bin appears to favor the double-Schechter model for both star-forming and quiescent SMFs.
Parameters of the best-fit single- and double-Schechter functions and corresponding BIC are shown in Table \ref{tab:params}.
An important subtlety, however, is that to robustly discriminate the necessity of a double-Schechter parameterization requires having measurements below the threshold at which the upturn begins to dominate ($\lesssim 10^{9.5}$ \msol) which, for quiescent galaxies in our dataset, lies near and/or below the stellar mass completeness limit.
Therefore, a direct comparison of the BIC for quiescent SMFs does not carry the same weight as for the total or star-forming SMFs, and more-so does not rule out the potential existence of such an upturn in all cases.

What else is immediately noticeable is that, similar to the full galaxy sample, the shape of both the star-forming and quiescent SMFs are strongly dependent on local environment, in agreement with some recent studies \citep{Vulcani2012, Davidzon2016}.
In Figure \ref{fig:schechter_params} we examine the shapes of the star-forming and quiescent SMFs by plotting likelihood contours of the single-Schechter parameters $M^*$ and $\alpha$\footnote{Note that in order to perform a self-consistent comparison we exclusively use the single-Schechter function.
}
.
These contours are derived from a series of Monte-Carlo simulations where for each iteration we resample the number counts of each SMF from a Poisson distribution and refit a single-Schechter function.
What we see is that the Schechter parameters appear to separate into two general loci for low- and high-density SMFs, where ``low'' and ``high'' in this context are defined as being below and above log(1+$\delta_{\mathrm{gal}}$) = 0.5 respectively.
This observation goes to suggest that the environmental footprint in the SMF is established in the intermediate densities of galaxy groups.

%Nevertheless, many past studies have found no significant environmental dependence in the SMF shape \citep{Peng2010, Bolzonella2010, Giodini2012, Vulcani2013, vanderBurg2013}, and in fact, an environment-independent SMF for star-forming galaxies is a key assumption in the quenching model introduced by \citet{Peng2010}.
Nevertheless, many past studies have found no significant environmental dependence in the SMF shape \citep{Peng2010, Bolzonella2010, Giodini2012, Vulcani2013, Calvi2013, vanderBurg2013}, and in fact, an environment-independent SMF for star-forming galaxies is a key assumption in the quenching model introduced by \citet{Peng2010}.
While it is difficult to robustly identify the cause of this discrepancy several important caveats are worth mentioning.
First, at these redshifts ($0.7 < z < 1.2$) the aforementioned studies have been limited to stellar masses $>10^{10}$ \msol .
However, if we restrict our comparison to $>10^{10}$ \msol\ we see only a mild reduction in the significance of this result (see bottom panels of Figure \ref{fig:schechter_params}).
It is important to note that for the star-forming SMF this observation is almost exclusively driven by a difference in the low-mass slope $\alpha$ (whereas for the quiescent SMF it is a combination of $M^*$ and $\alpha$).

The next set of caveats regard definitions, interpretations, and range of galaxy environment.
In general, environment can act on ``local'' and ``global'' scales and effects on observed galaxy properties can be sensitive to the scale probed by the chosen environmental metric \citep{Muldrew2012}.
The importance of the distinction between these environmental definitions in relation to the galaxy SMF are discussed in \citet{Vulcani2012} and \citet{Vulcani2013}.
As discussed in Section 3.3, the essence of the environmental metric used in this work is based on a Voronoi tessellation which has been shown to be one of the most effective methods for recovering the local density field \citep{Darvish2015}.
However, this metric has a complex interrelation with global environment \citep[e.g.][]{Weinmann2006, DeLucia2012, Hearin2016} which we do not attempt to reconcile here.
Therefore, caution should be exercised when comparing the results of this work to studies which employ global density metrics.
In addition, the range of environments probed is not always consistent as some studies utilize cosmological galaxy surveys whereas others utilize dedicated surveys of large scale structures.
The scarcity of massive large scale structures in cosmological surveys, such as zCOSMOS and VIPERS \citep{Bolzonella2010, Davidzon2016}, limits their ability to place constraints for the highest density environments.
The typical density contrast between low- and high-density environments in these studies is $\sim$10$\times$, roughly an order of magnitude less than the range probed in ORELSE.

Another distinction is that many of these past studies use only a single rest-frame colour to define star-forming subsamples.
This is important because star-forming galaxies enshrouded by dust can be easily misclassified as quiescent, and in particular, the fraction of star-forming galaxies that have a significant level of dust attenuation correlates with stellar mass \citep{Wild2014, Martis2016}.
Thus using a single rest-frame colour introduces a mass-dependent bias in the classification of star-forming and quiescent galaxies.

\subsection{A Semi-Empirical Model of the Galaxy Stellar Mass Function}
\label{sec:model}

In recent years there have been many studies devoted to modeling the evolution of the galaxy stellar mass function.
A large subset of these take the approach of using the observed SMF at high redshift as an initial condition and allowing it to evolve forward in time after incorporating a variety of known aspects of galaxy evolution such as growth due to star-formation, quenching, etc. \citep[e.g.][]{Leja2015, Tomczak2016, Contini2017, Steinhardt2017}.
Despite various differences in the exact approach taken, one common result from these studies is that there exists some tension between the observed evolution of the galaxy SMF and the well-known correlation between star-formation rate and stellar mass \citep[e.g.][]{Noeske2007}.
One emerging picture from many of these past studies is that galaxy-galaxy merging appears to be necessary at some level in order to help relieve this disagreement.
Furthermore, these past studies have been limited to modeling the SMF of field galaxies.
Here we present a simple semi-empirical model in a similar context as these past works but with the aim of explaining the SMF of dense environments.

Galaxy merging is well-recognized as a crucial mechanism in the mass assembly of galaxies \citep[e.g.][]{LeFevre2000, Guo2008, Lin2010, RodriguezGomez2016} as well as a mechanism for building the diffuse stellar component of the intra-cluster medium known as the intra-cluster light, or ICL \citep[e.g.][]{Willman2004, Rudick2006, Murante2007, Contini2014}.
Furthermore, the galaxy-galaxy merger rate has been shown to be elevated in overdense relative to underdense regions by factors of 3-4$\times$ \citep{Lin2010, Kampczyk2013}.
This increase is generally believed to be driven by galaxy group-like environments where moderate velocity dispersions (a few $\times$100 km/s) would not suppress the likelihood of galaxy mergers as could be the case for galaxy clusters with velocity dispersions $\gtrsim$1000 km/s.
In a case study of a galaxy proto-cluster at $z=1.62$ \citet{Lotz2013} estimate a merger rate that is 3-10$\times$ greater than a sample of field galaxies matched in redshift and stellar mass.
These results provide strong evidence that galaxy-galaxy mergers must play a crucial in shaping the galaxy SMF of large scale structures.

Given that the evolution of the galaxy stellar mass function at $z<5$ is predominantly in its normalization, whereas the shape remains roughly constant \citep[e.g.][]{Ilbert2013, Tomczak2014, Grazian2015, Mortlock2015} it is reasonable to utilize the shape of the field SMF at $z\sim0.8$ as measured in our data as a starting point for understanding the SMF in dense environments.
We begin by generating a sample of $\approx$10$^6$ galaxies with stellar masses down to $10^6$ \msol\ distributed according to the best-fit double-Schechter function of the SMF in our lowest density environments (-0.5 $<$ log(1+$\delta_{\mathrm{gal}}$) $<$ 0.5).
Assuming this sample is representative of the galaxy population at $z_{\rm{start}}=5$ the simulation progresses forward incrementally in 100 Myr intervals to $z_{\rm{final}}=0.8$.
At each time-step (1) galaxies grow via star-formation, provided they are not quenched, (2) a number of galaxy pairs are selected to be merged, and (3) a number of galaxies are selected to be quenched.
The only property of this model that we allow to vary is the the total number of mergers that occur between the $z_{\rm{start}}$ and $z_{\rm{final}}$, all other prescriptions/assumptions are fixed as described below.

%%%%%%%%%%%%%%%%%%%%%%%%%%%%%%%%%%%%%%
%%%  HOW SIMULATED GALXIES EVOLVE  %%%
%%%%%%%%%%%%%%%%%%%%%%%%%%%%%%%%%%%%%%

To track the stellar mass evolution of a simulated galaxy, we first define individual ``stellar mass particles''.
Each stellar mass particle is treated as a single stellar population (SSP) and is subject to mass loss over time due to stellar evolution.
For this we adopt the mass loss formula presented in Equation 16 of \citet{Moster2013} which is based on the mass evolution of a model SSP from \citet{Bruzual2003} with an assumed \citet{Chabrier2003} IMF.

A simulated galaxy is defined as a collection of stellar mass particles which effectively represent its constituent stellar populations, either formed in situ or ex situ.
At the onset of the simulation, galaxies are assigned a single ``seed'' stellar mass particle, representing the galaxy's total stellar mass at $z_{\rm{start}}$, with an age randomly selected between 0-1 Gyr.
This constraint ensures that no stellar mass particle is older than the universe at $z_{\rm{start}}$.

%%%%%%%%%%%%%%%%%%%%%%%%%%%%%%%%%%%%%
%%%  STAR-FORMATION IN THE MODEL  %%%
%%%%%%%%%%%%%%%%%%%%%%%%%%%%%%%%%%%%%

\subsubsection{Prescription for Star-Formation}

Star-formation rates of galaxies are assigned and evolve based on the SFR-M$_*$ relations presented in Equations 2 and 4 of \citet{Tomczak2016}.
The left panel of Figure \ref{fig:sfrmass_fquiescent_simulation} shows the range of the SFR-M$_*$ relation between the redshift limits of the model.
In general, for galaxies below $10^{10.5}$ \msol\ this prescription is very close to a power law relation of $\rm{SFR} \propto M_*^{1.09}$.

First, galaxies are assigned a SFR based on their stellar mass at $z_{\rm{start}}$.
With each time-step of the simulation, galaxies that are not quenched generate a new associated stellar mass particle with a formation mass equal to the galaxy's instantaneous SFR multiplied by the time interval of the simulation (100 Myr).
At the end of each time-step a galaxy's SFR is set to the value of the SFR-M$_*$ relation at the corresponding redshift and stellar mass.

%%%%%%%%%%%%%%%%%%%%%%%%%%%%%%%%
%%%  QUENCHING IN THE MODEL  %%%
%%%%%%%%%%%%%%%%%%%%%%%%%%%%%%%%

\subsubsection{Prescription for Quenching}

Quenching in our model simply entails setting a galaxy's star-formation rate to zero at a given time-step.
Once a galaxy is quenched it will remain quenched for the remainder of the simulation (i.e. rejuvenated star-formation is not incorporated).
To quench galaxies we implement the following procedure.
We enforce that the quiescent fraction vs. stellar mass (i.e. the relative number of quenched galaxies to total at fixed M$_*$) at the end of the simulation match that of the measured quiescent fraction as dictated by the best-fit Schechter functions to the SMFs in the highest density bin (1.5 $<$ log(1+$\delta_{\mathrm{gal}}$) $<$ 2) shown in Figure \ref{fig:mfs_vor_sfqu} and Table \ref{tab:params}.
In essence this treats the quiescent fraction vs. stellar mass measured from the ORELSE data as a boundary condition for the endpoint of the simulation ($z_{\rm{final}}$=0.8).
In practice, at each time-step of the simulation we examine the quiescent fraction in bins of stellar mass, which at the start of the simulation is 0\%\ as all galaxies begin as star-forming.
We then randomly select galaxies to be quenched in order to match the quenching rate defined by these boundary conditions.
In this way, the quiescent fraction of each stellar mass bin gradually increases linearly with time from 0\%\ at $z_{\rm{start}}$ to its final value at $z_{\rm{final}}$.
An illustration of this is shown in the right panel of Figure \ref{fig:sfrmass_fquiescent_simulation}.

We note that this procedure involves extrapolating the Schechter fits to the ORESLE mass functions far below their stellar mass completeness limits which may not accurately represent galaxies in the low mass regime.
At first glance, it may seem that this approach considers only stellar mass as relevant towards determining when a galaxy will quench (a.k.a. ``mass-quenching'').
However, by having the quiescent fraction measured in the {\it highest density bin} be the endpoint, ``environmental-quenching'' is effectively encoded into the quenching scheme, albeit in an indirect way.
Nevertheless, this will only be the case (for certain) above the limiting stellar mass of our data ($\gtrsim$10$^{9.7}$ \msol), whereas the true evolution of quiescent fractions at lower stellar masses remains unknown.
Recent studies in the Local Group have suggested that the environmental quenching efficiency dramatically increases by $\gtrsim$$5\times$ for satellites at $\lesssim$$10^8$ \msol , likely tied to ram-pressure stripping \citep{Fillingham2015, Fillingham2016}.
However, it is unclear if this trend extends to higher redshifts where hot gaseous halos become more rare.

%%%%%%%%%%%%%%%%%%%%%%%%%%%%%%
%%%  MERGING IN THE MODEL  %%%
%%%%%%%%%%%%%%%%%%%%%%%%%%%%%%

\subsubsection{Prescription for Merging}

The final major prescription in this model is the implementation of galaxy-galaxy merging.
As mentioned earlier, the only feature of this model that is varied systematically is the total number of galaxies that merge between $z_{\rm{start}}$ and $z_{\rm{final}}$.
This is denoted as the fraction of galaxies merged ($f_{\rm{merged}}$) over the full duration of the simulation relative to the total number from the start.
We allow $f_{\rm{merged}}$ to vary between 0\%\ (i.e. no merging) and 95\%\ (i.e. only 5\%\ of galaxies remaining at $z_{\rm{final}}$).
In Figure \ref{fig:merger_simulation} we show the resulting SMFs at $z_{\rm{final}}$ of the simulation for a range of values of $f_{\rm{merged}}$.

We impose a redshift dependence wherein the merger rate (number of mergers per unit time) evolves as $\sim$$(1 + z)^{2.7}$, in accordance with recent work by \citet{RodriguezGomez2015} based on the Illustris hydrodynamic simulation \citep{Genel2014}.
It is important to note, however, that at present there is no broadly accepted consensus regarding redshift evolution of the galaxy-galaxy merger rate where some studies find that it increases with redshift \citep[e.g.][]{Hopkins2010a, Hopkins2010b, Man2012} while others find it to be roughly constant with redshift \citep[e.g.][]{Guo2008, Williams2011}.

Additionally, we enforce that minor mergers occur at 3$\times$ the frequency of major mergers \citep{Lotz2011}.
Major  and minor mergers are defined as having stellar mass ratios ($\mu_* \equiv M_{\mathrm{satellite}} : M_{\mathrm{primary}}$) between the limits of [1:4$-$1:1] and [$\leq$1:4] respectively.
For each merger event the less massive galaxy is ``destroyed'', having its stellar mass particles transferred to the more massive galaxy.
During this transfer we assume that 30\% of the stellar mass of the less massive galaxy is stripped and becomes part of the ICL.
While this implementation is overly simplistic as the actual amount of stellar mass stripped through the relaxation process will depend on, e.g., merging timescales and orbital velocities, this average value has been argued to be a reasonable first-order assumption from numerical simulations \citep[][and references therein]{Somerville2008, Contini2014, Contini2017}.
We keep track of this fraction of the total stellar mass ($M_{*,\,\mathrm{ICL}} / M_{*,\,\mathrm{total}}$).

\subsubsection{Comparing the Model to Observations}

We compare the model to the measured SMFs from Figure \ref{fig:mfs_vor} and identify which value of $f_{\rm{merged}}$ best reproduces the data.
This is done by scaling the simulated to the measured SMFs and minimizing $\chi^2$ for each realization of $f_{\rm{merged}}$.
We then marginalize over all realizations of $f_{\rm{merged}}$ based on their minimized $\chi^2$.
Despite its simplicity, the model is able to reproduce the shape of the measured SMFs reasonably well as shown in Figure \ref{fig:merger_simulation_results}.

We find that the measured SMFs in the three highest overdensity bins strongly favor scenarios where large majorities of galaxies merge in the simulation.
This implies that galaxy-galaxy mergers are vitally important in shaping the galaxy stellar mass function of dense environments, consistent with other recent works \citep{Davidzon2016, Steinhardt2017}.
Furthermore, we observe that the best-fit value of $f_{\rm{merged}}$ jumps up rapidly to 79\%\ for the intermediate density bin, 0.5 < log(1+$\delta_{\mathrm{gal}}$) < 1.0, and increases more modestly to the two higher density bins (Figure \ref{fig:merger_simulation_results}).
The implication from this is that intermediate environments (e.g. galaxy groups) are likely where most mergers would be occurring.
This is qualitatively consistent with expectations as galaxy groups are believed to be the environment most conducive to galaxy-galaxy merging due to their moderate velocity dispersions, whereas velocities of typical cluster galaxies may act to suppress merging \citep{Lin2010}.

Our model yields values of $M_{*,\,\mathrm{ICL}} / M_{*,\,\mathrm{total}}$ of 2-5\%.
These numbers are lower than recent estimates of the luminosity and stellar mass fractions of the ICL at low redshifts from the literature which tend to range between 10-40\% \citep[e.g.][]{Willman2004, Rudick2006, Gonzalez2007, Murante2007, Sand2011, Contini2014, Mihos2017}.
Although these values are in tension with each other a few caveats bear mentioning.
First, these studies use a variety of approaches making it difficult to fully understand the underlying systematics.
Some examples of these methods for obtaining ICL fractions include inferences from N-body and/or hydrodynamical simulations, direct measurements of emission from the ICL, and inferences from the rate of hostless Type Ia supernovae in the ICM.
Furthermore, these studies pertain to low redshifts ($z < 0.15$; roughly 5 Gyr later in cosmic time than this work), which could in part explain why our values are lower if the ICL is still building at $z \lesssim 0.8$.
This is in fact consistent with simulations which show that ICL fractions can roughly double at $z \leq 1$ \citep{Willman2004, Rudick2006}.
Finally, it is important to point out that our model does not include tidal stripping of stellar mass not associated with mergers as a separate channel for contributing to the ICL.
This would indicate that the ICL fractions from our model are underpredictions given that \citet{Contini2014} argue that most of a galaxy cluster's ICL component is formed through the stripping/disruption of satellites as opposed to mass loss from mergers.

However, we would like to remind the reader that this model presents a simplified view of galaxy evolution and relies on various assumptions and extrapolations of empirical relations well below regimes where current observational data are able to probe.
For example, some features of the model where systematic uncertainties may lurk include:

\begin{enumerate}
\item[$\bullet$] using empirical SFR-M$_*$ relations to construct SFHs \\[-4mm]
\item[$\bullet$] choice of quenching mechanism(s) \\[-4mm]
\item[$\bullet$] adopting a redshift-dependent merger rate \\[-4mm]
\item[$\bullet$] lack of infall of new galaxies/groups \\[-4mm]
\item[$\bullet$] lack of tidal stripping not associated with mergers
\end{enumerate}

It may also be the case that the assumption of 30\% satellite stellar mass loss per merger is wrong, or that this quantity is dependent on the merger stellar mass ratio $\mu_*$.
Furthermore, galaxies with higher stellar masses will generally also reside in more massive dark matter halos with greater gravitational potentials.
This will inevitably influence merger rates in a way that gives preference to more massive central galaxies as shown in Figure 6 of \citet{RodriguezGomez2015}.
In a follow-up paper we plan to develop this model further and examine the results in more detail, as well as add the remaining 7 survey fields of ORELSE, virtually doubling our sample size.

\section{Summary and Conclusions}
\label{sec:summaryandconclusions}

In this work we explore and compare the galaxy stellar mass function across a wide range of local environments as probed by the ORELSE survey \citep{Lubin2009}.
Leveraging extensive photometric and spectroscopic coverage from eight ORELSE fields hosting massive large scale structures at $0.6 < z < 1.3$ we measure the SMF down to $10^9$ \msol .
Utilizing the large number of spectroscopically confirmed redshifts ($>4000$) we employ a Monte-Carlo Voronoi tessellation algorithm (described in section \ref{sec:voronoi_mc}) to estimate environmental density.
By discretizing in narrow slices of redshift we construct 3D environmental overdensity maps between $0.55 < z < 1.3$ for each of our 8 fields.
These overdensity maps detail a wide range of environments from those occupied by field galaxies, log(1+$\delta_{\mathrm{gal}}$) $\sim 0$, to the central cores of massive galaxy clusters, log(1+$\delta_{\mathrm{gal}}$) $> 1.5$.

We define five bins of overdensity of width 0.5 dex between $-0.5 \leq$ log(1+$\delta_{\mathrm{gal}}$) $\leq 2$ and construct the galaxy SMF from galaxies that fall into these corresponding environments.
The resulting stellar mass functions show a strong dependence on local environment.
More specifically this dependence is manifested as a smooth and continuous enhancement in the numbers of higher- to lower-mass galaxies towards denser regions.
This finding echoes results from several recent works which found, at varying levels of significance, similar behavior in the SMF at these redshifts \citep{Bolzonella2010, Vulcani2012, vanderBurg2013, Mortlock2015, Davidzon2016}.
We next repeat this analysis for galaxies separated into star-forming and quiescent subpopulations based on rest-frame $U - V$ and $V - J$ colours.
Interestingly we find the same relative enhancement of high- to low-mass galaxies as with the total galaxy population for both star-forming and quiescent galaxies, albeit at lower significance for the latter subsample.
While this effect has been observed in a couple of recent studies \citep{Vulcani2012, Davidzon2016}, it is at odds with several other studies which found no statistically significant environmental dependence in the star-forming and/or quiescent SMFs \citep{Peng2010, Bolzonella2010, Giodini2012, Vulcani2013, vanderBurg2013}.

To help draw a link between the SMFs of low and high density environments we devise a simple semi-empirical model to test the importance of galaxy-galaxy merging.
We generate a large sample of simulated galaxies ($\approx$10$^6$) having stellar masses down to $10^6$ \msol\ distributed according the best-fit double-Schechter function of our two lowest-density bins (Table \ref{tab:params}).
Assuming this sample represents the universe at $z=5$ we evolve it forward in time, stopping at $z=0.8$ (the median redshift of ORELSE).
Within this model we allow galaxies to (1) grow via star-formation in accordance with the observed SFR-M$_*$ relation, (2) quench at a rate that reproduces the observed quiescent fraction at $z=0.8$, and (3) merge at a rate that is proportional to $(1+z)^{2.7}$ (see Section \ref{sec:model} for a full description).
Merger pairs are selected randomly, although we enforce that minor mergers occur at 3$\times$ the rate of major mergers \citep{Lotz2011}.
Furthermore, with each merger event we assume that 30\%\ of the stellar material of the lower-mass galaxy is stripped and becomes part of the intra-cluster light \citep{Somerville2008, Contini2014}.

The only property of the simulation that we systematically vary is the overall number of galaxies that merge between the beginning and end ($0.8 < z < 5$), which we quantify as a fraction relative to the initial number of galaxies ($f_{\rm{merged}}$).
In general we find that the observed SMFs in dense environments (log(1+$\delta_{\mathrm{gal}}$) > 0.5) are strongly favored by versions of the model where large numbers of galaxies merge ($f_{\rm{merged}} \geq 79$\%).
Because a large value of $f_{\rm{merged}}$ is found for the intermediate density bin, 0.5 < log(1+$\delta_{\mathrm{gal}}$) < 1.0, we argue that most of the mergers necessary to transform the SMF would have to occur in intermediate density environments such as galaxy groups.
This is consistent with expectations as galaxy groups with moderate velocity dispersions are generally thought to be the most ideal environment for galaxy-galaxy mergers as opposed to galaxy clusters \citep{Lin2010}.
Nevertheless, the overall implication from our model is that galaxy-galaxy merging is a very relevant process for shaping the galaxy stellar mass function of dense environments, consistent with other recent studies \citep{Davidzon2016, Steinhardt2017}.

Another deduction that can be made from our model is the fraction of stellar mass locked in the ICL of galaxy clusters ($M_{*,\,\mathrm{ICL}} / M_{*,\,\mathrm{total}}$).
Because the only way to increase $M_{*,\,\mathrm{ICL}}$ in the model is through mass loss from mergers this fraction will, in general, monotonically increase with the total number of mergers ($f_{\rm{merged}}$).
The best-fit models yield ICL fractions around 2-5\%.
In general this is less than low-$z$ estimates from past literature which find a wide range of values mostly between 10-40\% \citep[e.g.][]{Willman2004, Rudick2006, Gonzalez2007, Murante2007, Sand2011, Contini2014, Mihos2017}.
An important caveat is that our model does not include tidal stripping of stellar mass {\it not associated with mergers}, which \citet{Contini2014} argue contributes more to the ICL than mass loss from mergers.
This suggests that our derived ICL fractions are underestimated.
However, because many of these studies pertain to low redshifts ($z < 0.15$; roughly 5 Gyr later in cosmic time than this work) the low values from the model presented in this paper could suggest that the ICL fraction continues to grow at $z \lesssim 0.8$.
In fact, simulations show that the ICL fraction can increase by as much as 2$\times$ between $z=1$ and $z=0$ \citep{Willman2004, Rudick2006} which, if true, would place our estimates at the lower end of measured values at $z\sim0$ even without the additive effects of tidal stripping.
Nevertheless, it is also important to note that estimates of ICL fractions come from a variety of different methodologies including direct measurements of the low surface brightness emission of the ICL, measurements of the rate of hostless Type Ia supernovae in the ICM, and various types of simulations (N-body, semi-analytic, hydrodynamical, etc.).
Each of these approaches have their own inherent sources of systematic uncertainty which can cloud comparisons between them.

It is important to note that this semi-empirical model presents a simplified picture of galaxy evolution and is subject to various systematic uncertainties.
Nevertheless, the main conclusions regarding the necessity of prevalent merging to reproduce the stellar mass distribution of cluster and group galaxies, and the preference of such merging to occur in group-scale environments, is almost certainly not sensitive to such uncertainties.
In a follow-up analysis we plan to incorporate the remaining ORESLE survey fields (for which data are still being processed presently), which will effectively double the number of galaxies in our sample.
This will help to further improve constraints on the SMF shape as a function of environment, as well as open up the possibility of subsampling by redshift.
Furthermore, we plan to develop/test the semi-empirical model further and analyze results from it in more detail.

\section*{Acknowledgements}

This material is based upon work supported by the National Aeronautics and Space Administration under NASA Grant Number NNX15AK92G.
Part of the work presented herein is supported by the National Science Foundation under Grant No. 1411943.
This research made use of Astropy, a community-developed core Python package for Astronomy (Astropy Collaboration, 2013).
A.R.T. would also like to thank Chris Fassnacht for aiding in the reduction of photometric data through helpful discussions and provision of well-documented code.
S.M. acknowledges financial support from the Institut Universitaire de France (IUF), of which she is senior member.
A portion of this work made use of the Peloton computing cluster operated by the Division of Mathematical and Physical Sciences at the University of California, Davis.
Work presented here is based in part on data collected at Subaru Telescope as well as archival data obtained from the SMOKA, which is operated by the Astronomy Data Center, National Astronomical Observatory of Japan.
This work is based in part on observations made with the Large Format Camera mounted on the 200-inch Hale Telescope at Palomar Observatory, owned and operated by the California Institute of Technology.
A subset of observations were obtained with WIRCam, a joint project of CFHT, Taiwan, Korea, Canada, France, at the Canada-France-Hawaii Telescope (CFHT) which is operated by the National Research Council (NRC) of Canada, the Institut National des Sciences de l'Univers of the Centre National de la Recherche Scientifique of France, and the University of Hawaii.
UKIRT is supported by NASA and operated under an agreement among the University of Hawaii, the University of Arizona, and Lockheed Martin Advanced Technology Center; operations are enabled through the cooperation of the East Asian Observatory.
When the data reported here were acquired, UKIRT was operated by the Joint Astronomy Centre on behalf of the Science and Technology Facilities Council of the U.K. 
This work is based in part on observations made with the Spitzer Space Telescope, which is operated by the Jet Propulsion Laboratory, California Institute of Technology under a contract with NASA.
Spectroscopic observations used in the work presented here were obtained at the W.M. Keck Observatory, which is operated as a scientific partnership among the California Institute of Technology, the University of California and the National Aeronautics and Space Administration.
The Observatory was made possible by the generous financial support of the W.M. Keck Foundation.
We wish to recognize and acknowledge the very significant cultural role and reverence that the summit of Mauna Kea has always had within the indigenous Hawaiian community.
We are most fortunate to have the opportunity to conduct observations from this mountain.

\nocite{*}
\bibliographystyle{mnras}

\end{document}